\documentclass{aa}
\usepackage{amsmath,amstext}
\usepackage{txfonts}
\usepackage{graphicx}	% Including figure files
\usepackage{amsmath}	% Advanced maths commands
\usepackage{amssymb}	% Extra maths symbols
\usepackage{soul}       % for strikeout, can be removed after comments are treated
\usepackage[T1]{fontenc}
\usepackage{natbib}
\bibpunct{(}{)}{;}{a}{}{,}
\newcommand{\spitzer}{{\it Spitzer}}

\newcommand{\Msun}{M\ensuremath{_\odot}}

\newcommand{\mdust}{\ensuremath{M_{\rm{dust}}}}
\newcommand{\tdust}{\ensuremath{T_{\rm{dust}}}}

\newcommand{\clacc}{classification accuracy}
\newcommand{\fracrsk}{R\ensuremath{\sigma^{\text{pred}}_{\star}}}%fraction of predictions with reliable s_y
\newcommand{\rsk}{\ensuremath{\sigma^{\text{pred}}_{\star}}}
\newcommand{\mocassin}{{\tt MOCASSIN}}
\newcommand{\smodels}{SN model SED}%or MSSD?
\newcommand{\smodelss}{SN model SEDs}
\newcommand{\sdust}{dust species}
\newcommand{\dout}{\bgroup \markoverwith{\rule[0.2ex]{0.1pt}{0.4pt}\rule[0.8ex]{0.1pt}
{0.4pt}}\ULon}

\usepackage{tikz,xcolor,hyperref}
\definecolor{regalia}{rgb}{0.32, 0.18, 0.5}
\definecolor{azure}{rgb}{0.0, 0.5, 1.0}
\definecolor{darkbyzantium}{rgb}{0.36, 0.22, 0.33}
\definecolor{darkslateblue}{rgb}{0.28, 0.24, 0.55}
\definecolor{persianindigo}{rgb}{0.2, 0.07, 0.48}
\definecolor{skobeloff}{rgb}{0.0, 0.48, 0.45}
\definecolor{darkcoral}{rgb}{0.8, 0.36, 0.27}
\definecolor{darklava}{rgb}{0.28, 0.24, 0.2}

\definecolor{lime}{HTML}{A6CE39}
\DeclareRobustCommand{\orcidicon}{
	\begin{tikzpicture}
	\draw[lime, fill=lime] (0,0) 
	circle [radius=0.16] 
	node[white] {{\fontfamily{qag}\selectfont \tiny ID}};
	\draw[white, fill=white] (-0.0625,0.095) 
	circle [radius=0.007];
	\end{tikzpicture}
	\hspace{-2mm}
}

\foreach \x in {A, ..., Z}{%
	\expandafter\xdef\csname orcid\x\endcsname{\noexpand\href{https://orcid.org/\csname orcidauthor\x\endcsname}{\noexpand\orcidicon}}
}

\begin{document}
\title{Inferring properties of dust in supernovae with neural networks}

\author{Zoe Ansari{\orcidA{}}\inst{1} \and
Christa Gall{\orcidB{}}\inst{1} \and
Roger Wesson{\orcidC{}}\inst{2} \and
Oswin Krause{\orcidD{}}\inst{3} 
}
 \authorrunning{Ansari, Gall, Krause, Wesson}

\institute{
DARK, Niels Bohr Institute, University of Copenhagen, Jagtvej 128, 2200 Copenhagen, Denmark
\and
Department of Physics and Astronomy, University College London, Gower Street, London WC1E 6BT, UK
\and
Department of Computer Science, University of Copenhagen, Universitetsparken 1, 2100 Copenhagen, Denmark
}
%------------------------------------------
%----------------------------------ABSTRACT----------------------------------
\abstract
%context
{Determining properties of dust that formed in and around supernovae from observations remains challenging. This may be due to either incomplete coverage of data in wavelength or time, but also due to often inconspicuous signatures of dust in the observed data.}
%aims
{Here we address this challenge using modern machine learning methods to determine the amount and temperature of dust as well as its composition from a large set of simulated data. We aim to quantify if such methods are suitable to infer quantities and properties of dust from future observations of supernovae.} 
%methods
{We developed a neural network consisting of eight fully connected layers and an output layer with specified activation functions that allowed us to predict the dust mass, temperature, and composition as well as their respective uncertainties for each single supernova of a large set of simulated supernova spectral energy distributions (SEDs). 
We produced the large set of supernova SEDs for a wide range of different supernovae and dust properties using the advanced, fully three-dimensional radiative transfer code \mocassin. We then convolved each SED with the entire suite of {\it James Web Space Telescope} (JWST) bandpass filters to synthesise a photometric data set. We split this data set into three subsets which were used to train, validate, and test the neural network.  
To find out how accurately the neural network can predict the dust mass, temperature, and composition from the simulated data, we considered three different scenarios. First, we adopted a uniform distance of $\sim$ 0.43 Mpc for all simulated SEDs. Next we uniformly distributed all simulated SEDs within a volume of 0.43--65 Mpc and, finally, we artificially added random noise corresponding to a photometric uncertainty of 0.1 mag. Lastly, we conducted a feature importance analysis via SHapley Additive exPlanations (SHAP) to find the minimum set of JWST bandpass filters required to predict the selected dust quantities with an accuracy that is comparable to standard methods in the literature.   
}
%results
{We find that our neural network performs best for the scenario in which all SEDs are at the same distance and for a minimum subset of seven JWST bandpass filters within a wavelength range 3--25 $\mu$m. This results in rather small root-mean-square errors (RMSEs) of $\sim$ 0.08 dex and $\sim$ 42\,K for the most reliable predicted dust masses and temperatures, respectively. For the scenario in which SEDs are distributed out to 65 Mpc and contain synthetic noise, the most reliable predicted dust masses and temperatures achieve an RMSE of $\sim$ 0.12 dex and $\sim$ 38\,K, respectively. Thus, in all scenarios, both predicted dust quantities have smaller predicted uncertainties compared to those in the literature achieved with common SED fitting methods of actual observations of supernovae. Moreover, our neural network can well distinguish between the different dust species included in our work, reaching a classification accuracy of up to 95\% for carbon and 99\% for silicate dust. 
}
%conclusions
{
Although we trained, validated, and tested our neural network entirely on simulated SEDs, our analysis shows that a suite of JWST bandpass filters containing NIRCam F070W, F140M, F356W and F480M as well as MIRI F560W, F770W, F1000W, F1130W, F1500W, and F1800W filters are likely the most important filters needed to derive the quantities and determine the properties of dust that formed in and around supernovae from future observations. We tested this on selected optical to infrared data of SN 1987A at 615 days past explosion and find good agreement with dust masses and temperatures inferred with standard fitting methods in the literature. 
}

%------------------------------------------
\maketitle
%------------------------------------------
%------------------------------------------
\section{Introduction}
\label{s:intro}

%BIG PICTURE
The origin of dust in galaxies in the Universe remains debated. Large amounts of dust are observed in galaxies and quasars in the early and local Universe \citep[e.g.][]{2003A&A...406L..55B,2003MNRAS.344L..74P,2010A&A...522A..15M,2015Natur.519..327W,2008ApJ...687..848W,2010ApJ...712..942M, 2018Natur.553...51M}, some of which require a rapid and efficient dust formation process \citep[e.g.][]{2007ApJ...662..927D,2011A&A...528A..14G,2011A&ARv..19...43G, 2012ApJ...756..164F}. 
There is growing evidence that core collapse supernovae (CCSNe), which mark the death of short-lived massive stars, are efficient dust producers likely responsible for the observed large amounts of dust in galaxies \citep{2011A&ARv..19...43G, 2014Natur.511..326G,2016MNRAS.463L.112F, 2018ApJ...868...62G, 2020MNRAS.496.3668D}. 
An alternative to the rapid in situ dust production in CCSNe is grain growth in cold molecular clouds in the interstellar medium \citep[ISM, e.g.][]{2009ASPC..414..453D} from rapidly produced dust grain seeds and heavy elements by CCSNe.

%SETTING THE STAGE
Dust masses inferred from observations of supernovae (SNe) range from less than about 10$^{-4}$ \Msun\ in young CCSNe of a few hundred days old to about 0.1--1.0 \Msun\ in old CCSN remnants of a few 100 -- 1\,000 years of age. From a handful of CCSNe that have observationally been monitored over several years, it is evident that the amount of dust gradually increases over about 25--30 years \citep{2011A&ARv..19...43G,2014Natur.511..326G,2015MNRAS.446.2089W,2016MNRAS.456.1269B, 2018ApJ...868...62G}. Observations of older supernova remnants (SNRs) such as Cas A \citep{2021MNRAS.504.2133N}, N49 \citep{2010A&A...518L.139O}, Sgr A East \citep[$\sim$ 0.02 \Msun\, and $\sim$ 10\,000 years old,][]{2015Sci...348..413L},
G11.2$-$0.3 ($\sim$ 0.34 \Msun), G21.5$-$0.9 ($\sim$ 0.29 \Msun), and G29.7$-$0.3 ($\sim$ 0.51 \Msun) \citep{2019MNRAS.483...70C} confirm that on average about $\sim$ 0.3 \Msun\ of CCSN produced dust is sustained over a period of about 3\,000 years.  
While this is sufficient to account for the total dust mass observed in local as well as high redshift galaxies \citep{2018ApJ...868...62G}, the final amount of dust released into the ISM may still depend on the efficiency of dust destruction and re-formation behind diverse short and long time-scale reverse shocks launched by the forward shock interaction with either the CSM \citep[e.g.][]{2012MNRAS.424.2659M, 2019MNRAS.482.1715M} or ISM \citep[e.g.][]{2012ApJ...748...12S, 2016A&A...590A..65M}.

Inferring dust quantities as well as properties from observations is challenging. Typically, the amount of dust and its temperature is determined by fitting the thermal dust emission in the near- to far-infrared (far-IR) wavelength range with dust models at different levels of complexity \citep[e.g.][]{2009ApJ...700..579R, 2011A&ARv..19...43G, 2015MNRAS.446.2089W, 2015ApJ...800...50M, 2019MNRAS.482.1715M, 2021arXiv210907942C}. However, the most common dust species are rather featureless in this wavelength range with silicates having the most prominent emission feature at around 10--12 micron \citep[][]{1984ApJ...285...89D, 2010ARA&A..48...21H}, which also could appear featureless for cold dust and/or dust with large grains. 
Due to limited computational power or insufficient data, the manifold of dust model parameters can often neither be fully explored nor constrained. This leads to dust mass estimates that may vary over an order of magnitude \citep[][]{2018ApJ...868...62G}. 

Warm and cold dust ($\lesssim$ 500\,K) in nearby SNe and SNRs has been detected in the mid- to far-IR wavelength range with telescopes such as WISE, SOFIA, ALMA or the {\it Herschel} mission (2009--2013) \citep[e.g.][and references therein]{2012ApJ...760...96G, 2014ApJ...782L...2I, 2019MNRAS.488..164D, 2011A&ARv..19...43G,  2018ApJ...868...62G}, and notably the \spitzer\ {\it Space Telescope}, which observed during its cold (2003--2009) and warm phase (2009--2020) about 380 CCSNe out of about 1100 SNe in total \citep[see for a summary][]{2019ApJS..241...38S}.    
The next telescope in line with the right sensitivity to observe dust that either is newly formed or heated and to possibly constrain some dust species will be the {\it James Web Space Telescope} \citep[JWST,][]{2006SSRv..123..485G}. With instruments onboard, such as the Near-Infrared Camera and Spectrograph (NIRCam, NIRSpec), the Near-Infrared Imager and Slitless Spectrograph (NIRISS), and the Mid-Infrared Instrument (MIRI) imaging as well as spectroscopic observations of CCSNe in the wavelength range 0.6 -- 28 $\mu$m will be possible. However, the wavelength range of JWST is shorter than the \spitzer\ Infrared Spectrograph wavelength range that extended out to $\sim$ 38 $\mu$m, thus JWST will preferentially allow to probe the hot and warm dust regime but will not be suitable to probe the cold dust regime at which the majority of the large dust masses in SNRs are detected.

In this paper, we investigated whether modern machine learning algorithms can be used to determine the dust mass, temperature, and possible grain species from the signatures dust imprints in the spectral energy distributions (SEDs) of SNe.
We trained a neural network to predict such dust quantities from a simulated set of SEDs of CCSNe with different dust quantities and properties. The SEDs were produced using the fully three-dimensional photoionisation and dust radiative transfer code \mocassin \footnote{\color{blue}https://mocassin.nebulousresearch.org} \citep{2003MNRAS.340.1136E, 2005MNRAS.362.1038E} exploring a large parameter space of dust and SN properties. Assuming that the SNe are distributed within maximally 65 Mpc, we then convolved the SEDs with the suite of available JWST NIRCam (0.6--5.0 $\mu$m) and MIRI (5.0--28 $\mu$m) bandpass filters to synthesise a photometric data set. The use of simulated data was essential for this work since unfortunately, the presently existing wealth of observational data of dust in and around SNe is insufficient.  

The neural network was optimised to predict the total dust mass, dust temperature, and dust species. The data input to the neural network included the entire photometric data set, which consists of 293\,236 SEDs and the redshift for each SED. 
To obtain a practical method, we performed a feature selection method to find the minimum number of JWST filters to estimate the dust properties. Furthermore, we trained the neural network to obtain an estimate on the uncertainties of the predicted quantities (i.e. dust mass, temperature and species). We then identified the most reliable predictions using self-defined and common performance evaluation metrics, which also provide information about the overall performance of the neural network.

%EXPLAIN THE SECTIONS
In Section~\ref{s:simdata} we describe the simulated data set which sets the basis of our analysis and which we used to train our machine learning algorithm, which is described in Section~\ref{s:NN}. In Section~\ref{s:evaluation} we describe the metrics that we employed to evaluate the performance of the neural network, and discuss possible caveats in Section~\ref{s:caveats}. We present our results in Section~\ref{s:results} and discuss the implications of our results on future observations and the SN dust community in Section~\ref{s:discussion}. We conclude in Section~\ref{s:conclusion}.  
Throughout the paper we assume a $\Lambda$CDM model
with $H_{0}= 70$ (km/s) Mpc, and $\Omega_{0}=0.3$ \citep{2017Natur.551...85A}. We applied the above mentioned assumptions on our simulated data set whenever needed, via a built-in library from 
{\tt astropy}\footnote{\color{blue}https://docs.astropy.org/en/stable/api/astropy.cosmology\\.FlatLambdaCDM.html\#astropy.cosmology.FlatLambdaCDM}.

%------------------------------------------
%------------------------------------------
\section{Simulated data}
\label{s:simdata}
Here, we describe the simulated data set, which consists of simulated SN SEDs from which we synthesised a photometric data set using the entire suite of JWST NIRCam and MIRI bandpass filters. 
We describe how we dealt with either exceptionally faint or bright sources with respect to the JWST detection / sensitivity limits. Furthermore, we define three different scenarios, in each of which we derived a different data set from the simulated data set to train the neural network and test its performance for predicting the SN dust quantities and properties.

\subsection{\mocassin}
\label{ss:mocassin}
\mocassin\ (Monte Carlo Simulations of Ionised Nebulae) is a fully three-dimensional radiative transfer code that propagates radiation packets using a Monte Carlo technique \citep{2003MNRAS.340.1136E,2005MNRAS.362.1038E}. % 
Arbitrary distributions of material can be represented within a Cartesian grid. The material can consist of gas, dust, or both. In each grid cell, the thermal equilibrium and ionisation balance equations are solved to determine the physical conditions. For dusty models, \mocassin\ uses standard Mie scattering theory to calculate the effective absorption and scattering efficiencies for a grain of radius $a$ at wavelength $\lambda$, from the optical constants of the material. Any type of grain size distribution and mixture of materials may be specified.

The material is illuminated by a radiation source or sources, which can be discrete point sources, or a diffuse source present within each grid cell. The spectral energy distribution of the illuminating source can be a simple blackbody (BB) or an arbitrary spectral shape such as a stellar atmosphere model. The radiation field is described by a composition of a discrete number of monochromatic packets of energy \citep{1985ApJ...288..679A} for all sources. At each location, the Monte Carlo estimator \citep{1999A&A...345..211L} derives the mean intensity of the radiation field. The contribution of each energy packet to the radiation field at each location is defined by its path through the grid.

To synthesise different SEDs of SNe with dusty shells (hereafter \smodelss) using \mocassin\
we defined a set of parameters for the underlying radiation source (the SN), the dust itself and its location. 
Specifically, in our simulation, our chosen radiation source is
a central blackbody, which is defined by a temperature ($T_{BB}$) and a luminosity ($L_{BB}$). The range of the two parameters follows typical measurements of SN photospheres up to a few hundred days past explosion. The range of radii used in our models covers both the expected radii of SNe ejecta up to $\sim$1\,000 days after explosion, as well as larger radii at which pre-existing dust flash-heated by a SN explosion could give rise to infrared emission. For the dust, we considered two prominent grain species, which are 
amorphous carbon and astronomical silicates with optical constants taken from \citet{1996MNRAS.282.1321Z} and  \citet{1984ApJ...285...89D}, respectively. For our simulations, we considered that all the dust consists of either 100\% carbon, 100\% silicates, or is a 50:50 mixture of the two dust species. The range of initial dust masses is limited to $10^{-5} - 10^{-1}$ \Msun. The upper dust mass limit is partly motivated by the long run-time of simulations of \smodelss{} have, if a lot of dust is present. Another reason is that the mean dust mass for SNe and SNRs is 0.4  $\pm$ 0.07 \Msun\ \citep{2018ApJ...868...62G}, but the dust temperatures for large dust masses ($\gtrsim$ 10$^{-2}$ \Msun) in some SNe is $\lesssim$ 50 K \citep{2014Natur.511..326G}. Even with JWST such cold dust will not be easily detected. 
Furthermore, we considered only single grain sizes ranging between 0.005 -- 5 $\mu m$. Typically, such grain sizes are present in for example the Milky Way \citep[e.g.]{1977ApJ...217..425M} and observed in some SNe \citep[e.g.][]{2014Natur.511..326G,2015MNRAS.446.2089W, 2020ApJ...894..111B}.
In total, our \smodelss{} are composed of seven parameters, for which we defined either a set of distinct choices or a range of values (some are described above). A summary of the entire parameter space is presented in Table~\ref{modelparameters}. To finally create our data set, each \smodels{} was synthesised from a set of parameters that was stochastically generated from this parameter space. This method ensures that the entire parameter space is uniformly exploited.

\begin{table}
\caption{Input parameters for the \mocassin\ models.}
\begin{tabular}{llll}
\hline
Parameter       & Value / Switch   & Unit & Description                         \\ 
\hline
T$_{\rm{BB}}$    & (4 -- 14) $\times$ 10$^{3}$ & K  & BB-temperature                    \\
L$_{\rm{BB}}$     & (1 -- 100) $\times$ 10$^4$ & L$_\odot$   & BB-luminosity    \\
R$_{\rm{out}}$   & (1 -- 20) $\times$ 10$^{16}$ & cm    & Outer  dust shell radius\\
R$_{\rm{in}}$    & 0.8 $\times$ R$_{\rm{out}}$ & cm & Inner dust shell radius      \\
M$_{\rm{d}}$    & 10$^{-5}$ -- 10$^{-1}$ & M$_\odot$ & \tablefootmark{a}Dust mass                      \\
$a$              & 0.005 -- 5   & $\mu$m & \tablefootmark{a}Grain size                \\

\hline
$\kappa ,X$   & silicates, carbon &  & Grain species\\
            & 50:50 mixture &  &\\
\hline
\end{tabular}
\tablefoot{
\tablefoottext{a}{ Values are chosen from a logarithmic distribution.}
}
\label{modelparameters}
\end{table}

In this work, we used \mocassin\ version 2.02.73 to synthesise 293\,236 model SEDs. 
We constructed a cubical Cartesian grid with 11 cells on each side of the 3D grid to model the dusty shells, which are defined by an inner and outer radius of the shell, R$_{\rm{in}}$ and R$_{\rm{out}}$, respectively. We modelled one-eighth of the grid cube (shell) with the illuminating source in one corner. Assuming spherical symmetry, this cube segment was then scaled to the full cube for an effective resolution of 21$^3$ cells \citep{2003MNRAS.340.1153E}.
We used 10$^6$ energy packets in most of our simulations. This relatively low number ($\sim$750 energy packets per grid cell) ensures that the \mocassin\ models run very quickly. However, at wavelengths where only a few photons are emitted, the SEDs are affected by small number of statistics and hence dominated by noise. Therefore, for \mocassin\ models with dust masses lower than $10^{-4}$ \Msun{} in which few photons are reprocessed to longer wavelengths, we used 10 times as many energy packets to reduce the statistical noise in the SEDs at longer wavelengths (e.g. 5-30 $\mu m$).

\begin{figure}
    \centering
    \includegraphics[width=0.5\textwidth]{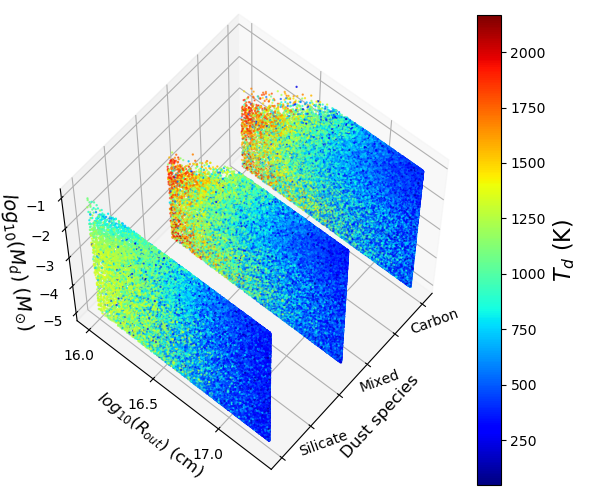}
    \caption{Coverage of \smodelss{} in \mdust, R$_{\rm{out}}$, and \sdust\ parameter space. The colour bar represents \tdust\ of the \smodelss, with blue, denoting the coldest (200\,K) and red, the hottest (2\,200\,K) temperatures. }
    \label{fig:high-runtimee}
\end{figure}

For efficiency reasons, we set a maximum run-time of two minutes for each model. For most regions of the investigated parameter space, the \mocassin\ models have a run-time of a few seconds, but models with both a small shell radius ($\lesssim$ 4$\times$10$^{16}$cm) and a high dust mass ($\gtrsim$ 10$^{-2}$M$_\odot$) have very high optical depths and thus, time out. This results in a slightly non-uniform filling of the entire parameter space. 
Furthermore, any dust grains in a simulation which reach the sublimation temperature of its species (1\,400\,K for silicate dust, 2\,200\,K for carbon dust) are considered to have evaporated and are not included when calculating the SED. The final dust mass is then either lower than the input dust mass or dust may even be no longer existing. Consequently, for mixed-chemistry models, the composition is altered from 50:50 to a higher carbon fraction due to the higher sublimation temperature of carbon dust. \mocassin\ does not directly provide the final dust mass and composition if dust evaporation occurs, but they are easily extracted from the output grid files by summing the dust masses in cells where the temperature is below the dust sublimation temperature. Some dust evaporation occurs in about 5\% of our models.
Figure~\ref{fig:high-runtimee} shows the final distribution of the SN model SEDs in the dust mass, temperature, and species \mocassin\ output-parameter space.

\subsubsection{Synthetic photometry of optical and mid-IR JWST bandpass filters}
\label{ss:synth_phot}

The JWST is equipped with two imaging cameras, NIRCam and MIRI. The two cameras have in total six narrow and 31 broad bandpass filters available that cover the wavelength ranges $0.6-5$~$\mu m$ (NIRCam) and $5-30$~$\mu m$ (MIRI).
As a next step in preparing the data set for our neural network, we convolved the \smodelss{} with both NIRCam and MIRI bandpass filters (hereafter filters) in order to synthesise a photometric data set. 
To do so, we used the python program 
{\tt Pyphot}\footnote{\href{https://mfouesneau.github.io/docs/pyphot/libcontent.html}{\color{blue}https://mfouesneau.github.io/docs/pyphot/libcontent.html}}. This program has a built-in library of transmission curves of different filters. Since {\tt Pyphot} also allows customised transmission curves, we imported transmission curves for NIRCam and MIRI filters from the Spanish virtual observatory\footnote{\href{http://svo2.cab.inta-csic.es/svo/theory/fps3/}{\color{blue}http://svo2.cab.inta-csic.es/svo/theory/fps3/}}. For each NIRCam and MIRI filter, we first calculated the integrated flux in units of Jansky via {\tt Pyphot}, which then were converted to AB magnitudes as
\begin{equation}
\label{eq:3}
    M_{AB}= 2.5 \times (23 - \log_{10}(F_{\nu}(\lambda_{\text{obs}})) -48.6,
\end{equation}
following the definition of \citet{2002astro.ph.10394H}.

\subsubsection{JWST detection limits}
\label{ss:jwstsenbri}

For the final step, we considered that our synthetic photometric data set contains magnitudes in some filters that would either be too bright or too faint to be detected with JWST.
In order to filter out data with magnitudes that practically cannot be observed (hereafter missing values),
we adopted the pre-calculated point-source continuum detection limits \citep{2015PASP..127..686G, 10.1117/1.JATIS.3.3.035001} that have been derived using the JWST exposure time calculator \citep[ETC,][]{2016SPIE.9910E..16P} 
for a signal-to-noise ratio (S/N) of 10 and exposure times of 21.4 s and 10\,000 s for the saturation and sensitivity limits, respectively.
A visualisation of these limits for all NIRCam and MIRI filters is shown in the appendix in Figures~\ref{fig:sen_bri_NIRCam} and ~\ref{fig:sen_bri_MIRI}, respectively. 

\subsection{Three scenarios}
\label{ss:secnarios}

Typically, CCSNe occur in different types of galaxies at different distances. Consequently, distant CCSNe appear fainter than the same nearby CCSNe because their brightness decreases with distance as
\begin{equation}
    \label{eq:1}
    F_{\nu}(\lambda_{\text{obs}})=\frac{L_{\nu}(\lambda_{\text{emit}})(1+z)}{4\pi D_{L}^{2}}
\end{equation}
with $F_{\nu}(\lambda_{\text{obs}})$ the observed flux as a function of the observed wavelength, $\lambda_{\text{obs}}$, in units of Jy; $L_{\nu}(\lambda_{\text{emit}})$ the emitted luminosity at the emitted wavelength, $\lambda_{\text{emit}}$; $D_{L}$ the luminosity distance and $z$ the redshift. The observed wavelength is given by $\lambda_{\text{obs}}=\lambda_{\text{emit}}(1+z)$.

This implies that the SEDs of CCSNe are  redshifted and a well-defined bandpass filter will sample the light from a bluer wavelength region of the intrinsic CCSN spectrum compared to the restframe wavelength range of the bandpass filter. In extreme cases (e.g. at high redshift) such an effect may cause a non-negligible degeneracy between dust properties and redshift. 
In what follows, we define three individual scenarios that are used to test if some quantities and properties of dust formed in and around CCSNe, such as the dust mass, \mdust\, dust temperature, \tdust\ and the dust species can be determined with neural networks.  

For the first scenario, S1, we simply assumed that all CCSNe are at the same, low redshift of $z$ = 0.0001, which corresponds to a distance of $\approx 0.43$ Mpc. For comparison, the distance of SN~1987A, the closest observed extragalactic CCSN is $\approx$ 0.49 $\pm$ 0.0009 (statistical) $\pm$ 0.0054 (systematic) Mpc \citep{2019Natur.567..200P} and the next closest CCSN, SN~1885A \citep{1989ApJ...341L..55F}, is $\sim$ 0.765 Mpc away.

Placing all \smodelss\ at the same such short distance has the advantage that the observed model magnitudes are nearly identical to the intrinsic magnitudes of \smodelss\ and thus, free of any possible degeneracy between dust properties and distance. Hence, we expect this scenario to be an ideal test case for the neural network. Moreover, from this scenario we can identify the smallest amount of dust detectable with JWST (see Section~\ref{ss:mocassin}). For simplicity reasons, here for S1, we only considered the upper sensitivity limit of the JWST filters but did not apply the lower saturation limits.

For the second scenario, S2, we assumed that all our simulated CCSNe are uniformly distributed within the redshift range 0.0001--0.015, which corresponds to a distance range of $\sim$0.43--65 Mpc. The decrease in brightness and shift in wavelength with increasing distance, 
together with the sensitivity and saturation limits of the JWST filters (see Figures~\ref{fig:sen_bri_NIRCam} and \ref{fig:sen_bri_MIRI}) place a limit on the distance out to which dust in CCSNe may be observed. Therefore, we chose $z$ = 0.015 (i.e. $\sim$ 65 Mpc) as an upper limit. This limit is based on the \smodelss, for which
the thermal dust emission of $10^{-5}$ \Msun\ carbon dust at a temperature of $\sim$ 2\,000\,K remains detectable (see Section~\ref{ss:jwstsenbri}) in at minimum 10 out of 28 NIRCam filters. 
 
The data sets of scenarios S1 and S2 solely consist of synthesised magnitudes of all available JWST filters without uncertainties. Therefore, as our third test scenario, S3, we used the data set of S2 and added synthetic photometric noise. 
We assumed that each synthesised magnitude is `observed' at S/N = 10, which translates into an uncertainty of 0.1 mag. 
This assumption is in line with what has been used to derive the detection limits (see Section~\ref{ss:jwstsenbri}). Hence, to create S3, we added randomly synthesised noise to the data of S2 as $m_{i, \rm{S3}} = m_{i, \rm{S2}} + \mathcal{N}(0, 0.1)$, with $m_{i}$ the magnitude of each JWST filters, and $\mathcal{N}(0, 0.1)$ as a randomly generated number taken from a Gaussian distribution with zero mean and $\sigma$ = 0.1.

%------------------------------------------
\section{Neural networks}
\label{s:NN}
Our analysis is based on training a deep neural network using simulated data (see Section~\ref{s:simdata}). The goal is to predict three dust quantities and properties, \tdust, \mdust\ and \sdust, 
together with a prediction of their respective uncertainties.
To conform with machine learning nomenclature, we refer to the set of photometric data that is synthesised from each \smodels{} using JWST filters, along with the redshift of the \smodels, as the input features.
We also refer to each \smodels\ that corresponds to each set of synthesised magnitudes, as a data point, since it is defined as a point in the input features' space.

In the following subsections, we describe the artificial neural networks and the corresponding hyperparameters. We also describe the specific type of neural network that we used and its corresponding optimal set of hyperparameters as well as a pre-processing method to treat the missing values in our data set. Furthermore, we describe the training process of our neural network, in which we defined target values for three dust quantities and properties. Thereafter, we explain an iterative feature selection procedure which we used to find the minimum set of the most important JWST filters, with which the dust quantities and properties can still be predicted with an acceptable accuracy.

\subsection{Artificial neural networks}
\label{ss:ANN}
An artificial neural network or in short, neural network, is a set of algorithms that is used to recognise relationships in a data set, and to find patterns. The structure of a neural network is inspired by biological neurons, and thus it mimics the methodology that biological neurons use to send signals to one another. Neural networks consist of one or more layers, known as hidden layers, between an input and an output layer. Each layer contains a set of neurons. The process of training a neural network consists of transferring information from the input layer to the output layer via a set of connections. Each connection is defined between each neuron in one layer to each neuron in the next layer. There are different methods to connect neurons and to transfer information between them. In the classic framework, each neuron of a given layer is connected to all neurons in the next layer. Layers that follow this pattern are called fully connected layers.
Another method for connecting neurons consists of convolutional layers, in which each neuron from a layer is only connected to a well defined set of neurons from the next layer. A neural network can be built using either one or a combination of different layers and different patterns.
To transfer the information, each layer applies an activation function to a set of weights associated with a set of neurons in the layer.

The output vector of each layer is defined as follows:
\begin{equation}
    \label{eq:layervalue}
    \mathbf{a}_{i}^l = \mathcal{H}^l \left(\sum_{j=1}^{p}\textup{W}_{j,i}^{l} \cdot \mathbf{a}_{j}^{l-1} + \mathbf{b}^{l-1} \right)
\end{equation}
where $\mathbf{a}^{l-1}$ is the input vector to layer $l$, $\textup{W}_{i,j}^{l}$ is a matrix that contains a set of weights from neuron $j$ in layer $l-1$ to neuron $i$ in layer $l$, $p$ is the number of neurons in the layer $l-1$, $\mathbf{b}^{l-1}$ is a vector of constant values assigned to neurons of layer $l$, known as thresholds, and $\mathcal{H}^{l}$ is an activation function for layer $l$. For the input layer (i.e. $l$=0) $\mathbf{a}_{i}^{l}= \mathbf{x}$, where $\mathbf{x}$ is the input feature vector for the neural network. 

The weights and the thresholds of neural networks are the model parameters that a neural network aims to optimise by improving its performance of estimating the target values.
In a forward-propagation process of a neural network training, the prediction error is first calculated using random weights. The prediction errors are quantified by a `loss function'.
In a subsequent back-propagation process \citep[e.g.][]{1986Natur.323..533R}, the weights are adjusted with the aim of minimising the loss. 
As the name suggests, the forward-propagation method iterates from the input via the hidden to the output layer, while the back-propagation is converse. 
This combination of forward- and back-propagation takes place within one epoch of training (hereafter epoch). Typically, several epochs are required to minimise the loss function and to improve the performance of the neural network.

Since the loss function can be non-convex, and finding a global minimum of a general non-convex function is NP-hard \citep{Murty1987SomeNP}, a neural network can be considered optimised when the loss function is converged to a `good' local minimum. To do so, minimisation algorithms, such as the classical gradient descent, are employed. The basic principle of such algorithms is to calculate the gradient of the loss function and step by step move in the direction as specified by the gradient, with the step size termed as the learning rate. 

Choosing the right learning rate is important as 
for a high learning rate the calculated loss with updated model parameters can jump over the local minimum, therefore can not converge to it. On the other hand, using a low learning rate, the algorithm takes a long time to reach the local minimum of the loss function. 

The batch gradient descent is a gradient descent optimisation method in which the neural network updates the weights only once per epoch for the entire training data set. Although this process is a fast approach for finding the local minimum of the loss function, the memory requirement for such computational task is large. A remedy to this is to employ a mini-batch gradient descent, which allows the neural network in each epoch to update the weights for a sub-sample of the data set separately. This subsample is called mini-batch, and the size of it is defined by the size of the mini-batch.

The classical gradient descent uses a fixed learning rate for the entire process. Since this is not optimal, other types of optimisation algorithms that can adjust the learning rate, such as Adaptive Moment Estimation \citep[ADAM,][]{2014arXiv1412.6980K} may be used instead.

Neural network parameters, such as the number of either hidden layers, neurons or epochs, the learning rate, the optimiser, the activation function for each layer and the size of the mini-batch, are referred to as hyperparameters. The hyperparameters affect the efficiency and performance of the neural network and, like the model parameters, need to be optimised to reach the best possible network performance. 
While the process of training a neural network adjusts the model parameters, usually
the hyperparameters must be manually fine-tuned for each science case and data set in question \citep[]{LeCun1998, Bengio2012, 2017arXiv170803888Y, 10.1145/3219819.3220058, 2020arXiv200707588W}.

\subsection{Our neural network}
\label{ss:ourNN}
We designed a neural network to estimate a set of target values, along with their uncertainties.
Our neural network aims to approximate a distribution for each target value with a given input feature, $\mathbf{x}$, of each data point and three target values correspond to three dust properties, $y^{\text{sim}}_{\mdust}$, $y^{\text{sim}}_{\tdust}$, and $y^{\text{sim}}_{\kappa}$. The neural network implements this approximation by maximising the log-likelihood of the target values under the assumption that the deviations follow a normal distribution, by approximating the mean ($m_{k}$) and standard deviation ($\sigma^{\text{pred}}_{k}$), which is the expected squared difference between the $y^{\text{pred}}$ and $y^{\text{sim}}$, as follows:
\begin{equation}
\label{eq:loss}
        \log\mathcal{L}(y^{\text{sim}},\mathbf{x})\ =\ \sum\limits_{n=1}^{N} \sum\limits_{k=1}^{K} \log \left(  \mathcal{N}(y^{\text{sim}}_{k,n}|m_{k}(\mathbf{x}_n),\sigma^{\text{pred}}_{k}(\mathbf{x}_n)) \right) \enspace
\end{equation}
where $N$ is the number of data points in the data set, while $K$ represents the number of target values. Therefore, each target value $y^{\text{sim}}_{k}$ is estimated by a mean $m_{k}$ (hereafter $y^{\text{pred}}_{k}$), and a standard deviation $\sigma^{\text{pred}}_{k}$, that represents the estimated uncertainty of $y^{\text{pred}}_{k}$.

\subsection{Hyperparameter tuning}
\label{ss:hyperparam}

To find the optimal set of hyperparameters for our neural network, we first explored combinations of 3--12 convolutional and fully connected layers. Each layer can have either four, 16, 32, 64, 128, 256, or 512 neurons. We used the standard Rectified Linear Units \citep[ReLU,][]{Maas13rectifiernonlinearities} and Parametric Rectified Linear Units \citep[PReLU,][]{2015arXiv150201852H} as non-linear activation functions between the input and the hidden layers. For the output layer, we used a linear activation function to predict the mean of the target values and an exponential linear unit  \citep[ELU,][]{2015arXiv151107289C} as activation function to predict the standard deviations of the target values. Using ELU as the activation function ensures that the estimated standard deviations are positive. 

We used six different learning rates of $10^{-6}$, $5\times 10^{-6}$, $10^{-5}$, $5 \times 10^{-5}$, $10^{-4}$, and $10^{-3}$ for the ADAM optimiser \citep{2014arXiv1412.6980K} to search for the local minimum of the loss function with mini-batch sizes of 32 and 64 data points. By comparing the validation and training loss of the neural network with different sets of hyperparameters, we found that the optimal set of hyperparameters consists of eight fully connected layers (one input and seven hidden layers) with 512, 256, 128, 64, 32, 16, eight, and four neurons in the first to the eighth layer, respectively. Furthermore, ReLU activation functions are best used between the layers together with a learning rate of $10^{-5}$ for the ADAM optimiser with mini-batch size of 64 data points. The number of epochs is chosen to be 2\,000 in S1 and 1500 for S2 and S3, in which the training and validation loss are converged.

\subsection{Missing data}
\label{ss:MV}
Considering the sensitivity and saturation limits (see Section~\ref{ss:secnarios}) for both MIRI and NIRCam filters, some \smodelss{} are not detectable in all filters over the entire wavelength range.
For instance, particularly bright or faint SEDs (or parts of the SEDs) result in magnitudes that either exceed the sensitivity limit or remain below the saturation limits of some filters.  
In reality, such cases would not lead to detections (magnitude measurements) and hence, may be considered as `missing values'. 
Here, for each filter, we replaced the synthesised magnitudes that fall outside the saturation and sensitivity limits with the magnitude of the sensitivity and saturation limits, respectively. This approach was inspired by the forced photometry measurement that is commonly used to study transients, for example for Pan-STAARS1\footnote{\href{https://outerspace.stsci.edu/display/PANSTARRS/PS1+Forced+photometry+of+sources}{\color{blue}https://outerspace.stsci.edu/display/PANSTARRS/ \\ PS1+Forced+photometry+of+sources}}. In this method, when a source is detected in a filter at a specific location in the sky, photometric values are forced to be extracted in other filters. These forced photometric values are either the actual magnitudes of the source, or the magnitude limits. 
%the limit magnitudes of the filters (i.e. the sensitivity or saturation limits).

\subsection{Neural network training preparation}
\label{ss:NNT}
To train our neural network with the set of hyperparameters that are defined in Section~\ref{ss:hyperparam}, we created a `training - validation - test' split from each of the data sets that are described in Section~\ref{ss:secnarios}. Particularly, out of a total of 293\,236 data points, we used 70\% (193\,536) as training, 15\% (49\,850) as validation and the remaining 15\% as test data set. 

We normalise
$y^{\text{sim}}_{M_{dust}}$ and $y^{\text{sim}}_{T_{dust}}$
% ) 
of all \smodelss{} as
\[
\large{
g(y^{\text{sim}})= \frac{y^{\text{sim}}}{y^{\text{sim,~max}}}
}
\]
with $y^{\text{sim,~max}}_{\mdust}=$ 0.1 $M_{\odot}$, and $y^{\text{sim,~max}}_{\tdust}=$ 2\,200\,K.
Moreover, we define a conditional function in which we arbitrarily assign each dust species (e.g. carbon, and silicate) a target value:
\[
\large{
g_{\kappa}(y^{\text{sim}})= \begin{cases}
    1,& \text{if~~} y^{\text{sim}} = \text{silicate}\\
    0.75, & \text{if~~} y^{\text{sim}} = \text{a mix of carbon and silicate}\\
    0.5, & \text{if~~} y^{\text{sim}} = \text{carbon}\enspace.
\end{cases}}
\]

We find that inferring the dust properties from \smodelss{} that contain no dust or only very small amounts of dust at cooler temperatures (\mdust\ < $5\times 10^{-5} M_{\odot}$ and \tdust\ < 800\,K) using neural networks is challenging (see Section~\ref{s:caveats} for further explanation). Therefore, to let the neural network differentiate between these \smodelss{} and \smodelss{} that contain recognisable dust, we defined a dedicated target value for this group of `no-dust' data points as $y^{\text{sim}}_{\text{no-dust}}$ = -0.5.

\subsection{Feature selection}
\label{ss:featselect}
The SHapley Additive exPlanations \citep[SHAP;][]{2017arXiv170507874L} is a framework that uses an additive feature attribution method to evaluate the importance of a certain input feature on the prediction of a neural network. In this framework, the Shapley values \citep{Shapley+2016+307+318} are calculated for each input feature based on cooperative game theory \citep{10.2307/1906951}. In this theory, to calculate the contribution of each input feature to a model's output,
the average marginal effect of feature $i$ is measured for all possible coalitions,
which represents the effect of feature $i$ on the model's output.
In an additive feature attribution method, for an input feature's vector $\mathbf{x}$, for a model $f$, a simplified local input feature's vector $\mathbf{x'}$, is defined for an explanatory model $\mathcal{F}$. The simplified local input feature's vector is a discrete binary vector, $\mathbf{x'}\in \{0,1\}^d$ (where d is the number of the input features), which means that either features are included or excluded. The explanatory model $\mathcal{F}$ is defined as
\[
    \mathcal{F}(x')= \phi_{0} + \sum_{i=1}^d \phi_{i} \mathbf{x'}_{i}\enspace,
\]
where $\phi_{0}$ is the base value of the model in the absence of any information, that is defined by the average of the model's output, and $\phi_{i}$ is the explained effect of feature $i$, known as the attribution of feature $i$. The $\phi_{i}$ shows how much feature $i$ changes the output of the model. The second term of the model $\mathcal{F}$ is the average over marginal contributions of each feature, over all possible coalitions. The absolute value of $\mathcal{F}(\mathbf{x'}_{n,i})-\mathcal{F}(\mathbf{x'}_{n,-i})$ indicates the importance of the feature $i$, where $\mathbf{x'}_{n,-i}$ represents that feature $i$ is not included in the input feature vector $\mathbf{x'}_{n,i}$.
Therefore, the Shapley values are defined as:
\begin{equation}
\label{eq:fi_1}
    \phi_{n,i} = \sum
    % _{{S \in \{0,1\}^d}\atop{z_i=1}} 
    \frac{|S|! (d - |S| - 1)!}{d!} 
    (\mathcal{F}(\mathbf{x'}_{n,i})
     - \mathcal{F}(\mathbf{x'}_{n,-i}))
\end{equation}
when the summation is over all feature subsets $ S\subseteq d$.

To calculate the Shapley values, all coalition values for all possible feature permutations must be sampled. Since the relation between the number of features and the number of possible feature permutations is exponential, for a large set of features the number of calculations in $\mathcal{F}$ is immense, and practically not feasible to implement. Therefore, the SHAP framework uses a fast approximation, Deep Learning Important FeaTures \citep[DeepLIFT,][]{2016arXiv160501713S,2017arXiv170402685S}, in which a linear approximation of Taylor series is used to approximate
\[
\mathcal{F}(\mathbf{x'}_{n,i}) - \mathcal{F}(\mathbf{x'}_{n,-i}) \enspace{,}
\]
in which the expectation values, $E[\mathbf{x'}]$, are calculated for all features and are used as referenced values in the input features vector, when the feature is omitted during the calculations.

Since the variance of the expectation values for N data points is roughly $ 1/\sqrt{N}$, using approximately 1\,000 data points gives an acceptable estimation for expectation values\footnote{\href{https://shap-lrjball.readthedocs.io/en/latest/generated/shap.DeepExplainer.html}{{\color {blue} https://shap-lrjball.readthedocs.io/en/latest/ \\
generated/shap.DeepExplainer.html}}}. Therefore, in this work, for each of our three test scenarios S1, S2 and S3, we used a sample of 5\,000 data points that we randomly chose from the training data set to approximate the expectation values for all features (i.e. $E[\mathbf{x'}_{i}]$ for $\forall i; i \in \mathbf{x'}$). Thereafter, we computed the Shapley values for 1\,000 randomly selected data points from the validation data set (see  Section~\ref{app:comp_cost} for the details of the computational cost). We selected the subsamples from the validation and training data sets with a random seed that we changed for each step in the feature selection process.
Therefore, we calculated the importance of each feature (i.e. filter) with index $i$ via
\begin{equation}
\label{eq:fi_3}
\phi_i = \frac 1 N\sum_{n=1}^{N} \left\lvert \phi_{n,i}\right\rvert \enspace,
\end{equation}
where N=1\,000. In each step, we removed the three filters that achieved the three lowest absolute Shapley values. Subsequently, in the next step, we trained the neural network using the reduced set of filters as the input feature's vector of the entire training data set and repeated the procedure. Considering that in each step, we removed the three least important filters, we performed the process for a total of 11 steps. Therefore, we are left with four filters out of 37 filters at the end of the process.

\begin{table*}
\caption{Comparison of neural network performance for estimating \mdust\ and \tdust\ in different scenarios for 4 different cases.
In case-1 and case-2, the training data set contains the preferred subset of JWST filters. In case-3 and case-4, the data set with the minimum subset of JWST filters is used to train our neural network.
In case-1, and case-3 the evaluation metrics are applied on the entire test data set. In case-2, and case-4 the evaluation metrics are applied on the predictions of the test data set that have \rsk.}
\centering
\label{tab:mdtd}
\begin{tabular}{p{1cm} p{1.5cm}|p{1.5cm} p{2cm} p{2.5cm}| p{1.5cm} p{2cm} p{2.5cm}}          % centered columns
\hline\hline                        
& &   &\mdust~(\Msun) & & &\tdust\ (K) &

\end{tabular}
\\
\begin{tabular}{p{1cm} p{1.5cm}| p{1.5cm} p{2cm} p{2.5cm}| p{1.5cm} p{2cm} p{2.5cm}} 
% \hline
Case & Scenario & bias (dex) & RMSE (dex) & 3$\sigma$ outliers (\%) & bias & RMSE & 3$\sigma$ outliers (\%)
\\
\hline

 &S1\tablefootmark{a} & -0.0084 & 0.1696 & 2.08 & 1.16 &18.16 &1.64 \\
 1&S2\tablefootmark{b} &-0.0786 & 0.3170 & 1.72 & 4.28& 31.25  & 2.05\\
 &S3\tablefootmark{c}& -0.0135 & 0.3120 & 7.01 & 6.31 & 59.90 & 2.02\\
 \hline
 &S1\tablefootmark{a} & -0.0110 & 0.0541 & 12.80 & 1.61 & 14.14 & 1.45 \\
 2&S2\tablefootmark{b}&  -0.0653 &0.1056 & 15.00 & 4.79 & 18.85 & 2.13\\
 &S3\tablefootmark{c}& -0.0355 & 0.1136 & 16.22  & 4.60 & 30.12 & 1.56\\
 \hline
 & S1\tablefootmark{d}& 0.0130 & 0.2075 & 8.19 & 1.39  & 56.52  & 1.92   \\
 3&S2\tablefootmark{e} & 0.0537 & 0.3985 & 4.10 & -0.62 & 37.92 & 2.01\\
 &S3\tablefootmark{f} & 0.0325 & 0.5522 & 2.25 & -2.69 & 78.55 & 2.34\\
 \hline
 & S1\tablefootmark{d} & 0.0013 & 0.0847 & 8.78  & 2.77 & 42.65  &1.41 \\
 4 &S2\tablefootmark{e} & 0.1018 & 0.1328 & 21.54 & -1.95 & 20.80 & 1.72\\
 &S3\tablefootmark{f} & -0.0235 & 0.1257 & 15.72 & -8.05 & 38.52 & 2.03\\
   
\hline                                            
\end{tabular}
\tablefoot{
{With the subset of JWST filters that are selected via the feature selection procedure as follows:}\\
\tablefoottext{a}{ NIRCam:$F070W$, $F115W$, $F140M$, $F150W$, $F210M$, $F300M$, $F335M$, $F360M$, $F430M$, $F444W$, $F460M$, $F466N$ , and MIRI: $F560W$, $F770W$, $F1000W$, $F1130W$, $F1280W$, $F1500W$, $F1800W$, $F2100W$, $F2550W$ }\\
\tablefoottext{b}{ NIRCam:$F070W$, $F115W$, $F140M$, $F150W$, $F182M$, $F187N$, $F200W$,$F250M$, $F277W$, $F300M$, $F322W2$, $F356W$, $F360M$,$F405N$ and MIRI: $F560W$, $F770W$, $F1000W$, $F1130W$, $F1280W$, $F1500W$, $F1800W$, $F2100W$}\\
\tablefoottext{c}{ NIRCam: $F070W$, $F115W$, $F140M$, $F356W$, $F460M$, $F480M$, and MIRI: $F560W$, $F770W$, $F1000W$, $F1130W$, $F1280W$, $F1500W$, $F1800W$}\\
\tablefoottext{d}{ NIRCam: $F460M$, and MIRI: $F560W$, $F770W$, $F1130W$, $F1280W$, $F1500W$, $F2100W$}\\
\tablefoottext{e}{ NIRCam: $F140M$, $F150W$, $F200W$, $F300M$, $F356W$, $F360M$, $F410M$, and MIRI: $F560W$, $F770W$, $F1000W$, $F1280W$, $F1500W$, $F1800W$}\\
\tablefoottext{f}{ NIRCam: $F070W$, $F140M$, $F356W$, $F480M$, and MIRI: $F560W$, $F770W$, $F1000W$, $F1130W$, $F1500W$, $F1800W$}}
\\
\end{table*}

\begin{table*}
\caption{Comparison of neural network performance for classifying dust species, and the fraction of predictions of the test data set that have \rsk\ to all the predictions from the test data set (\fracrsk), in different scenarios for four different cases. The definition of cases are the same as in Table~\ref{tab:mdtd}}
\centering
\label{tab:clacc}
\begin{tabular}{p{1cm} p{1.5cm} |p{0.9cm} p{5cm} p{0.1cm} |p{2cm}}
\hline \hline
& &  &\clacc\ rate (\%) &&\\
\end{tabular}
\\
\begin{tabular}{p{1cm} p{1.5cm} |p{2cm} p{2cm} p{2cm} |p{2cm}}
Case & Scenario & Carbon dust & Mixed dust  & Silicate dust &\fracrsk (\%)\\
\end{tabular}
\\
\begin{tabular}{p{1cm} p{1.5cm} |p{2cm} p{2cm} p{2cm} |p{2cm}}
\hline

&S1 \tablefootmark{a} & 86 & 95 &100 & -\\
% & 87 & 97 & 99& -\\
1& S2\tablefootmark{b} & 74 & 85 & 99 & -\\
% & 68 & 77 & 99& -\\
&S3\tablefootmark{c} & 72 & 75 & 98 & - \\
\hline
&S1 \tablefootmark{a} & 97 & 99 & 100 & 68\\
% & 96 & 99 & 99 & 77\\
2& S2 \tablefootmark{b} & 98 & 99 & 100 & 31\\
% & 97 & 98 & 100 & 24\\
&S3\tablefootmark{c} & 97 & 94 & 100 & 12\\
\hline
&S1 \tablefootmark{d} & 89 & 90 & 98 & -\\
3&S2\tablefootmark{e} & 73 & 79& 98 & -\\
&S3\tablefootmark{f} &  61& 63& 96 & -\\
\hline
&S1\tablefootmark{d} & 98 &100 & 100 & 48\\
4&S2\tablefootmark{e} & 94& 95 & 99& 34\\
&S3\tablefootmark{f} & 95 &57 & 99 & 07\\
\hline

\end{tabular}
\tablefoot{
{With the subset of JWST filters that are selected via the feature selection procedure as follows:}\\
\tablefoottext{a}-\tablefoottext{f}{: See the definitions in Table~\ref{tab:mdtd}.}}
\end{table*}

%-------------------------------
\section{Evaluation}
\label{s:evaluation}

In this section we describe the chosen evaluation metrics to evaluate the performance of our trained neural network. We address 
how we interpreted the resulting predictions for the dust species and 
how we treated the no-dust models in the performance evaluation.
Furthermore, we define criteria to estimate the reliability of the predictions via the predicted standard deviations as the outputs of the neural network. Finally we describe %4
the metrics for comparing the performance of the neural network in different steps of the feature selection process.

The performance evaluation of the predicted target values, $y^{\text{pred}}_{\mdust}$, $y^{\text{pred}}_{\tdust}$ and $y^{\text{pred}}_{\kappa}$, is applied on test data sets, and consists of three individual methods: root-mean-square error (RMSE), bias, and 3$\sigma$ outliers.
For the dust temperature, the residual of data point, $n$, is defined as $\Delta y_{\tdust,n}=y_{\tdust,n}^{\text{pred}} - y_{\tdust,n}^{\text{sim}}$. For the dust mass, due to the logarithmic distribution of \mdust\ in the simulated data set, we define the residual as $\Delta y_{\mdust,n}=log_{10}(y_{\mdust,n}^{\text{pred}}/y_{\mdust,n}^{\text{sim}})$ .
For both \mdust\ and \tdust\ the bias is defined as the mean of the residuals as
\[
\frac{1}{N}\sum\limits_{n=1}^{N} \Delta y_{n} \enspace,
\]
and the RMSE is defined as 
\[
\sqrt{\frac{1}{N}\sum\limits_{n=1}^{N} (\Delta y_{n})^{2}},
\]
where $n$ represents each data point, and $N$ is the number of data points in the test data set. 
Furthermore, for \mdust\ and \tdust\ we define the 3$\sigma$-outliers
% 3$\sigma$ outlier
as the predictions with $|\Delta y_{n}| > 3~ \times$ RMSE. 
%2
Moreover, due to the numeric representation of all \sdust\ (see Section~\ref{ss:NNT}) that are fed to the neural network, numeric target values are predicted. In order to interpret these numeric target values, we define each dust species as a `class'. This way,  we have the following classes: silicate, mixed, carbon and no-dust that we define by a conditional function as
\[
\large{
\mathcal{G}_{\kappa}(y^{\text{pred}})= \begin{cases}
    \text{silicate},& \text{if } y^{\text{pred}} > 0.875 \\
    \text{mixed}, & \text{if } 0.675 <y^{\text{pred}} \leq 0.875\\
    \text{carbon}, & \text{if } 0.25 < y^{\text{pred}} \leq 0.675 \\
    \text{no-dust}, & \text{if } y^{\text{pred}} \leq 0.25\enspace.
\end{cases}}
\]
Furthermore, to evaluate how well the neural network predicts the \sdust, we used the definition of true and false positives, and true and false negatives \citep[e.g.][]{journals/prl/Fawcett06} to build a confusion matrix. The classification accuracy for each dust species class is defined as the fraction of correct predictions out of the total number of predictions of each class from the neural network.

Moreover, we investigated whether the predicted uncertainties can be used to filter out uncertain predictions reliably. For this, we assumed that errors in the predicted quantities are approximately normal distributed with mean and variance as predicted by our model. Then, given a chosen confidence level the central confidence interval of the predicted quantity is 
\[
    |y_{k,n}-y^{\text{pred}}_{k,n}|\leq a1\times \sigma^{\text{pred}}_{k,n}\enspace,
\]
where $y^{\text{pred}}_{k,n}$ and $\sigma^{\text{pred}}_{k,n}$ are the predicted mean and standard deviation of the $k$th target value for the $n$th datapoint and $y_{k,n}$ is the unknown true value. The factor $a1$ is a parameter that depends on the chosen confidence level, where values $a1 = 1,2,3$ give rise to the 68\%, 95\%, and 99.7\% confidence levels, respectively.

With this, we define a threshold for the acceptable relative error, a2 and accept a predicted mean value as (likely) accurate if the width of the confidence interval is small compared to $y^{\text{pred}}_{k,n}$. For \mdust\ and \tdust\ this yields the criterion
% \[
\begin{equation}
\label{eq:criterion1}
    \frac {a1 \times\ \sigma^{\text{pred}}_{k,n}}{|y^{\text{pred}}_{k,n}|} < a2 \enspace,
\end{equation}
% \]
and for dust species, we use
\begin{equation}
\label{eq:criterion2}
    a1 \times\ \sigma^{\text{pred}}_{\kappa,n} < a2\enspace.
\end{equation}
In the following, we use $a2=0.2$ and $a1=1$. If a prediction satisfies equations~\ref{eq:criterion1} and~\ref{eq:criterion2}, we say that it has a reliable standard deviation(\rsk).
To compare the performance of the neural network in each step of the feature selection process, we used two values from the neural network output; i) the values that are reached by the loss function (i.e. Equation~\ref{eq:loss}), for the training and validation data sets at the end of the training process, ii) the ratio of the number of predictions that have \rsk, to the total number of predictions of the test data set (hereafter \fracrsk). 

Since we chose a fixed set of hyperparameters (see Section~\ref{ss:hyperparam}), for instance, a fixed number of epochs, the minimum loss achieved by the neural network in the training process in each step of the feature selection process can differ from the `absolute or true' minimum that could be achieved, if the hyperparameters were to be re-adjusted for each step. This is independent of the chosen subset of JWST filters and happens in all scenarios. Ideally, in order to reach the absolute minimum loss possible one should re-adjust the hyperparameters for each step. However, this is a very time consuming process. Additionally, this would make the entire feature selection process dependent on the training data set as well as on the subset of the JWST filters, while not providing further relevant information for all the steps necessary to obtain the final preferred subset of JWST filters.     

%------------------------------------------
\section{Caveats}
\label{s:caveats}

Typically, very low amounts of dust (less than about 10$^{-5}$ \Msun) are not easily observable in SNe, since the thermal dust emission is rather weak at the expected wavelengths. This means that in some of our \smodelss\ that contain such low amounts of dust, the thermal dust emission in the simulated SEDs may either not be clearly discernible from the emission of the SN or generally remains below the detection capabilities of JWST. Such \smodelss\ that exhibit barely noticeable or no dust signatures may therefore also remain largely unrecognised by our neural network.

\begin{figure}
    \centering
    \includegraphics[width=0.5\textwidth]{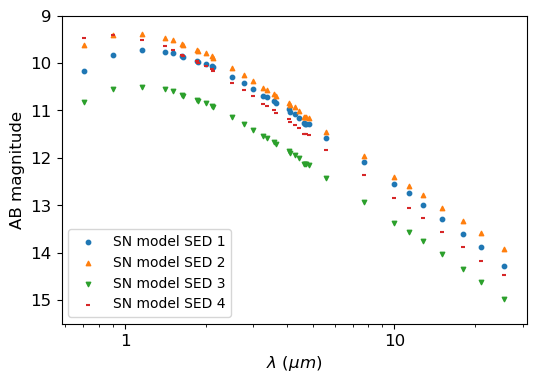}
    \caption{Example set of SN model SEDs with unrecognisable dust signatures. Each symbol represents a synthesised magnitude by a JWST filter and is shown at its central wavelength. The filled circles and triangles correspond to the SN model SED 1 and 2, that contain silicate dust, while filled dash and downward triangles correspond to SN model SED 3 and 4, respectively, both contain a mix of silicate and carbon dust. The amount and the temperature of dust in the SN model SEDs one to four, are about $2 \times 10^{-5}$ \Msun, 314\,K, $5 \times 10^{-5}$ \Msun, 329\,K, $ 10^{-5}$ \Msun, 620\,K, $ 10^{-5}$ \Msun, 617\,K  respectively.}
    \label{fig:no-dust}
\end{figure}

In what follows, we trained the neural network on the synthesised photometric data set for S1 to identify the \smodelss\ in this data set that have the lowest \mdust\ and \tdust\ that still can be recognised by the neural network. 
We find that for \smodelss\ with \mdust\ $ < 5\times 10^{-5}$ \Msun\ and \tdust\ $ < 800$ K the predicted dust properties have very large uncertainties (i.e. $|\Delta y_{\tdust}| \gtrsim 0.5 \times y_{\tdust}^{\text{sim}}$), causing so called catastrophic outliers. Consequently,
we trained the neural network again, but this time to label such \smodelss\  as `no-dust' data points (See Section~\ref{ss:NNT}), similar to the \smodelss\ that indeed contain no dust.
Figure~\ref{fig:no-dust} presents an example set of such no-dust \smodelss\ with \mdust\ and \tdust\ below the aforementioned thresholds. 
Due to the fact that from the no-dust \smodelss\ the predicted dust properties including their uncertainties are highly unreliable, we did not include these models in subsequent performance evaluations of the dust properties.

%------------------------------------------

\section{Results}
\label{s:results}

We investigated whether a neural network can be used as an effective tool to determine 
different properties of dust that formed in and around CCSNe from its spectral energy distribution. Since the number of observed SNe is too sparse to be used for such an endeavour, we simulated a total of 293\,236 SN SEDs (referred to as SN model SEDs), each with different dust properties. Then, we convolved each SN model SED with the entire suite of JWST NIRCam $+$ MIRI banpass filters (see details in Section~\ref{s:simdata}) to synthesise a photometric data set that is suitable for machine learning purposes.

For a step by step analysis we considered three different scenarios, which are described in more detail in Section~\ref{ss:secnarios}. In short, for the first scenario, S1, all  \smodelss\ are placed at the same, low redshift, $z = 0.0001$. In the second scenario, S2, we uniformly distributed the \smodelss\ within the redshift range 0.0001--0.015. In the third scenario, S3, we used the data set of S2 and added random noise that corresponds to a photometric uncertainty of 0.1 mag (see the details in Section~\ref{ss:jwstsenbri}). 
Comparing the outcome of these scenarios allowed us to examine how strongly the performances of the neural network and the feature importance change for our simulated data that are equipped with properties of real observations. 

\begin{figure}
    \centering
    \includegraphics[width=0.45\textwidth]{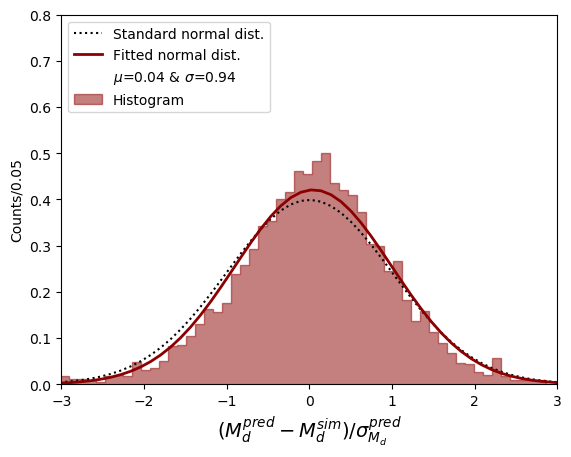}\\
    \includegraphics[width=0.45\textwidth]{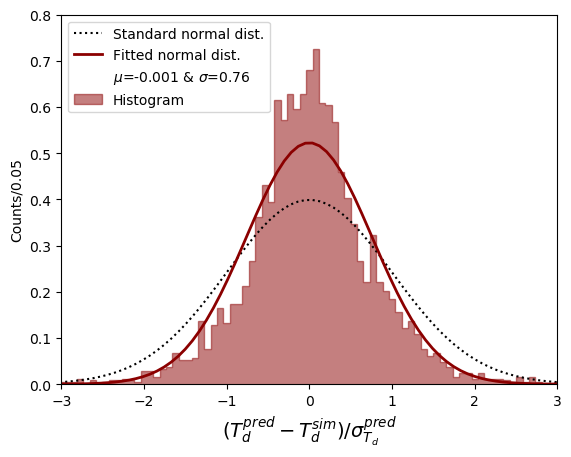}\\
    
    \caption{Comparison of the distribution of normalised prediction errors to a standard normal distribution. The histograms represent the distributions of {\large${(y^{\text{pred}} - y^{\text{sim}})}/{\sigma^{\text{pred}}}$} for \mdust\ ({\it top panel}), and \tdust\ ({\it bottom panel}), for a test data set predicted by the trained neural network with the entire suite of JWST filters, in S3. The dotted curves represent the standard normal distributions (i.e. $\mathcal{N}(0,1)$). The solid curves are the normal distributions fitted to each of the histograms with $\mu$=0.04 and $\sigma$=0.94 for \mdust, and $\mu$=-0.001 and $\sigma$=0.76 for \tdust.}
    \label{fig:cal}
\end{figure}

In our approach, we trained our neural network to predict the distribution of dust quantities given the SN model SEDs, $p(y^{\text{sim}}|\mathbf{x})\approx \mathcal{N}(y^{\text{sim}};y^{\text{pred}}, \sigma^{\text{pred}})$. To evaluate how well our estimated uncertainties align with the prediction errors, we analysed the distribution of the normalised prediction errors {\large${(y^{\text{pred}} - y^{\text{sim}})}/{\sigma^{\text{pred}}}$}. Under perfect neural network modelling circumstances, the distribution of these normalised values must follow a standard normal distribution.
Figure~\ref{fig:cal} shows histograms of the normalised
prediction errors for \mdust\ and \tdust\ of a test data set predicted by the trained neural network with the entire set of JWST filters, in S3, excluding the predictions that the neural network classifies them as no-dust.
By fitting a normal probability distribution function to the normalised prediction errors, we find that for \mdust\ distribution, a mean of 0.04, and a standard deviation of 0.94 are inferred. The inferred values for \tdust\ are the mean and standard deviation of -0.001 and 0.76, respectively. Therefore, the inferred standard deviations corresponding to \mdust\ and \tdust\ are 6\% and 24\% lower than for a standard normal distribution. This might indicate that the predicted uncertainties, $\sigma^{\text{pred}}$, are overestimating the prediction errors.

For each scenario, S1, S2, and S3 we discuss four cases of a performance evaluation. 
For case-1 and case-2 we evaluated the performances of our neural network that is trained on data sets that consist of preferred subsets of JWST filters (see Section~\ref{s:discussion} for further discussions on the selection of preferred subsets).
For case-3, and case-4 we evaluate the performances of our neural network that is trained with data sets that are constructed with a minimum subset of JWST filters (see definition section~\ref{s:evaluation}), with which the different dust quantities are predicted with an acceptable level of accuracy. The latter means that the fraction of reliable predictions, out of the entire test data set, is $\gtrsim 5\%$.
Furthermore, for case-1 and case-3 we apply the evaluation metrics on the entire test data set. For case-2 and case-4, we apply the metrics only on the subsample of the test data set that satisfies the criteria for being reliable predictions as defined in Section~\ref{s:evaluation}.

Tables~\ref{tab:mdtd} and~\ref{tab:clacc} summarise the outcome of the case by case performance evaluations of our neural network to predict \mdust, \tdust\ and to classify the \sdust\ for all three scenarios S1, S2 and S3.
Out of all scenarios and all cases we find that in S1 and for case-2, the RMSE of both \mdust\ and \tdust\ is the smallest and \fracrsk\ is maximal.
For case-2, the RMSE of \mdust\ increases from $\sim$ 0.05 dex in S1, to $\sim$ 0.1 dex in S2 and to $\sim$ 0.11 dex in S3. However, for \tdust\ the RMSE increases from about 14\,K in S1, only to $\sim$ 18\,K in S2. From S1 to S3, the RMSE of \tdust\ increases to $\sim$ 30\, K in S3. From both S1 to S3, in case-2, the fraction of $3\sigma$ outliers for \mdust\ target values increases.

The bias of the \tdust\ predictions for most of the scenarios for case-3 and case-4 is negative. This indicates that the neural network underestimates the \tdust\ target values (i.e. $y^{\text{pred}}_{\tdust}$). For case-1 and case-2 the bias is positive for \tdust\
in all scenarios. This indicate that the neural network overestimates the \tdust\ target values.
For instance, in case-3 for S1, the bias of 0.013 (dex) for \mdust\ represents that the average of \mdust\ estimations over all the test data set is about $10^{0.013} \approx 1.03$ times more than the simulated \mdust. For \tdust\ the average of \tdust\ estimations over all the test data set is about 1\,K more than the simulated \tdust\ values.

%cl-acc
As shown in Table~\ref{tab:clacc}, the highest classification accuracy for \sdust\ is achieved for S2 for case-2. For this, we find a classification accuracy of 97\%, 98\%, and 100\% for carbon, mixed, and silicate dust, respectively. Comparing the classification accuracy for each dust species, we find that for all scenarios and all cases, silicate dust is predicted with the highest accuracy. Carbon dust is predicted least accurately in all scenarios and cases, except in S3 for case-4. There, the \smodelss\ that are labelled as mixed dust are predicted with the lowest accuracy (57\%). In case-4 and S3, 42\% of the mixed dust species are predicted as carbon dust.

\begin{figure*}
    \centering

    \includegraphics[width=1\textwidth]{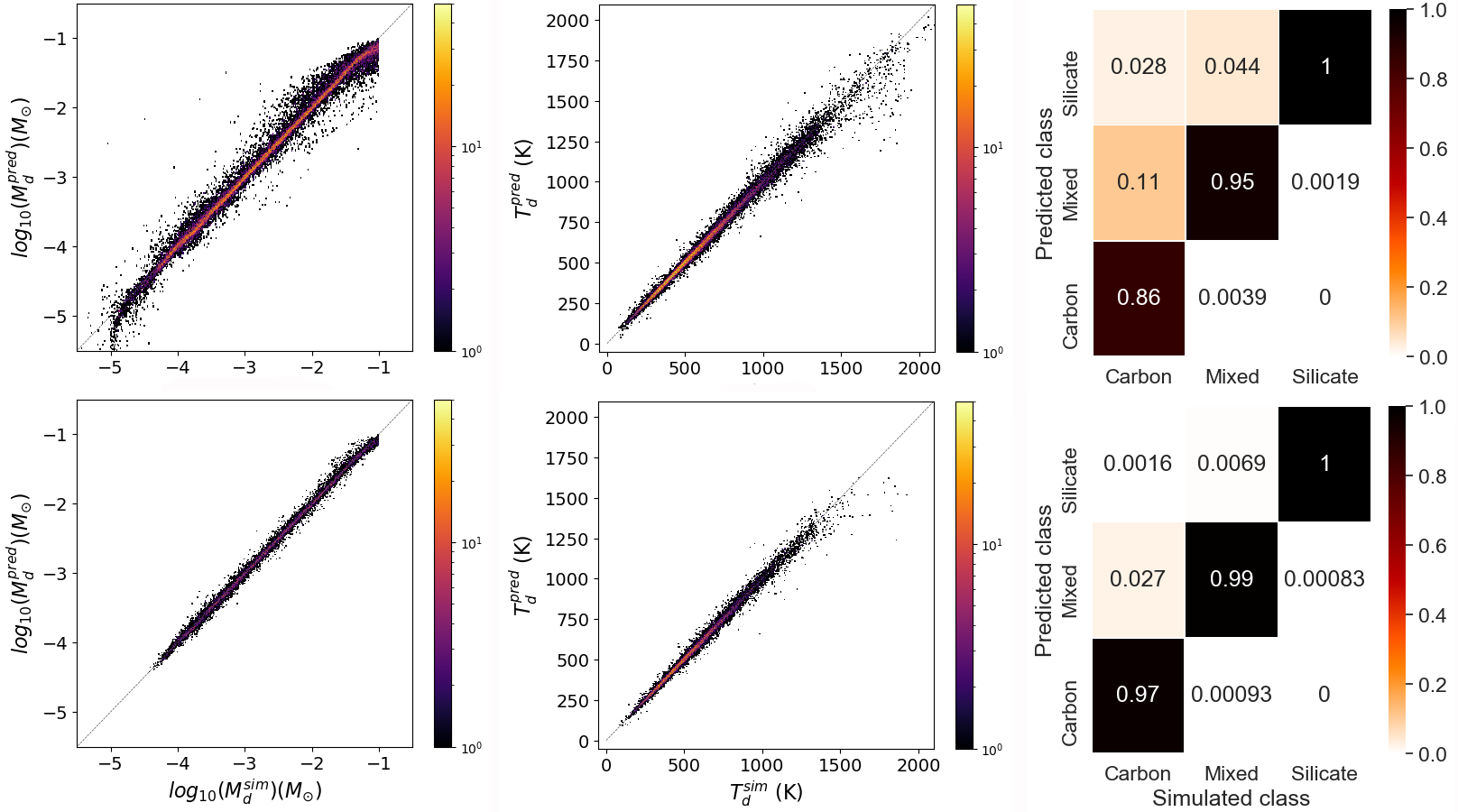}\\
    
    \caption{Performance of the neural network with the preferred subset of JWST filters, for S1. The \mdust\ ({\it left column}), and \tdust\ estimates ({\it middle column}), and \sdust\ classification ({\it right column}), are shown for all the predictions of the test data set ({\it top panel}), and the reliable predictions of the test data set ({\it bottom panel}). The \sdust\ classifications are shown in the format of confusion matrices that represent the simulated \sdust\ against the predicted \sdust. 
    The colour bars in the left and middle diagrams indicate the number of predictions, ranging from 1 (black) to 50 (yellow). The dashed lines mark where the predicted and  simulated values of \mdust\ and \tdust are equal. }
    \label{fig:1st_opt_test}
\end{figure*}

\begin{figure*}
    % \centering

    \includegraphics[width=1\textwidth]{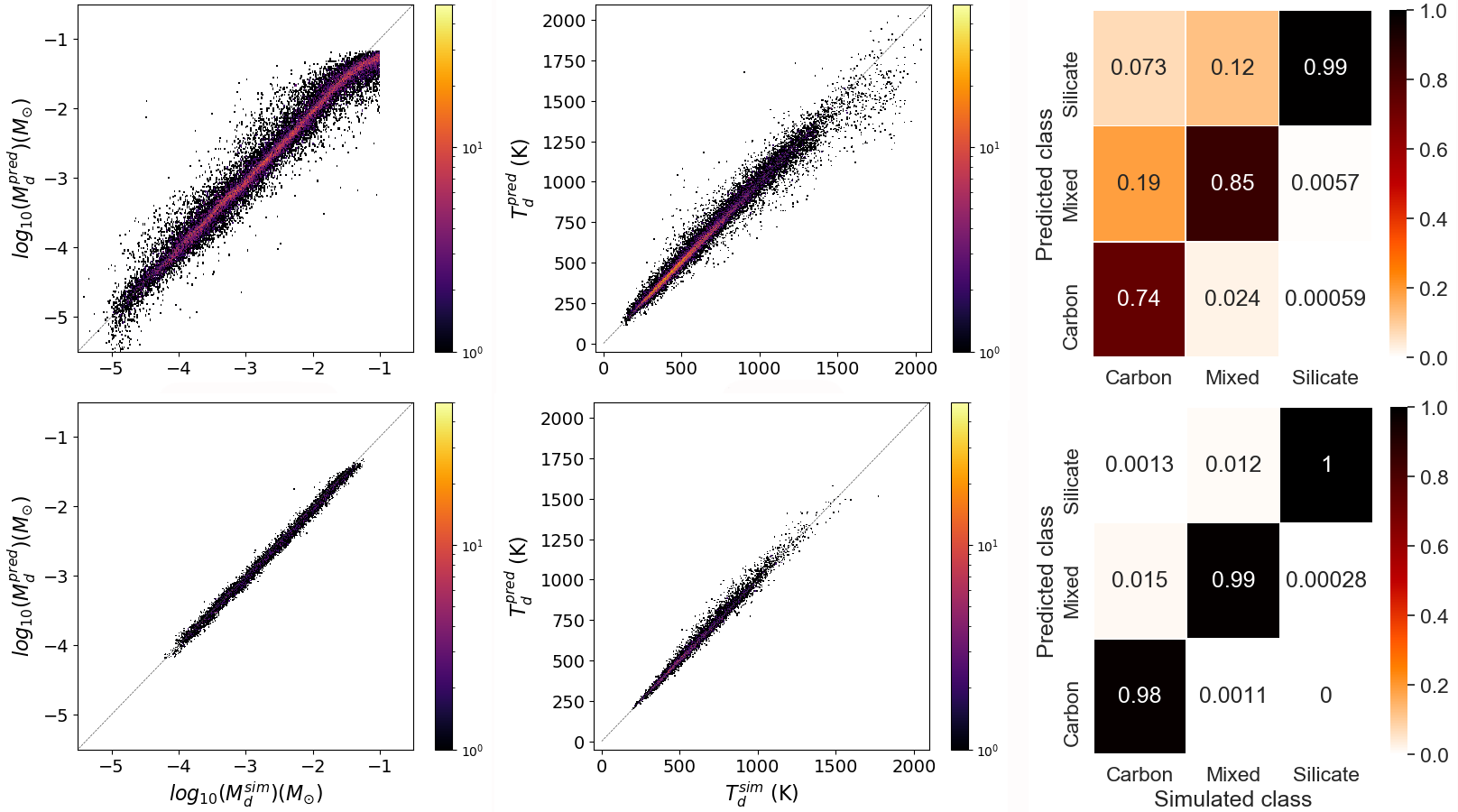}
    \caption{Performance of the neural network with the preferred subset of JWST filters, for S2. The definition of the panels, the variables, the dashed lines and the colour bars are the same as in Figure~\ref{fig:1st_opt_test}.}
    \label{fig:2nd_opt_test}
\end{figure*}

\begin{figure*}
    % \centering

    \includegraphics[width=1\textwidth]{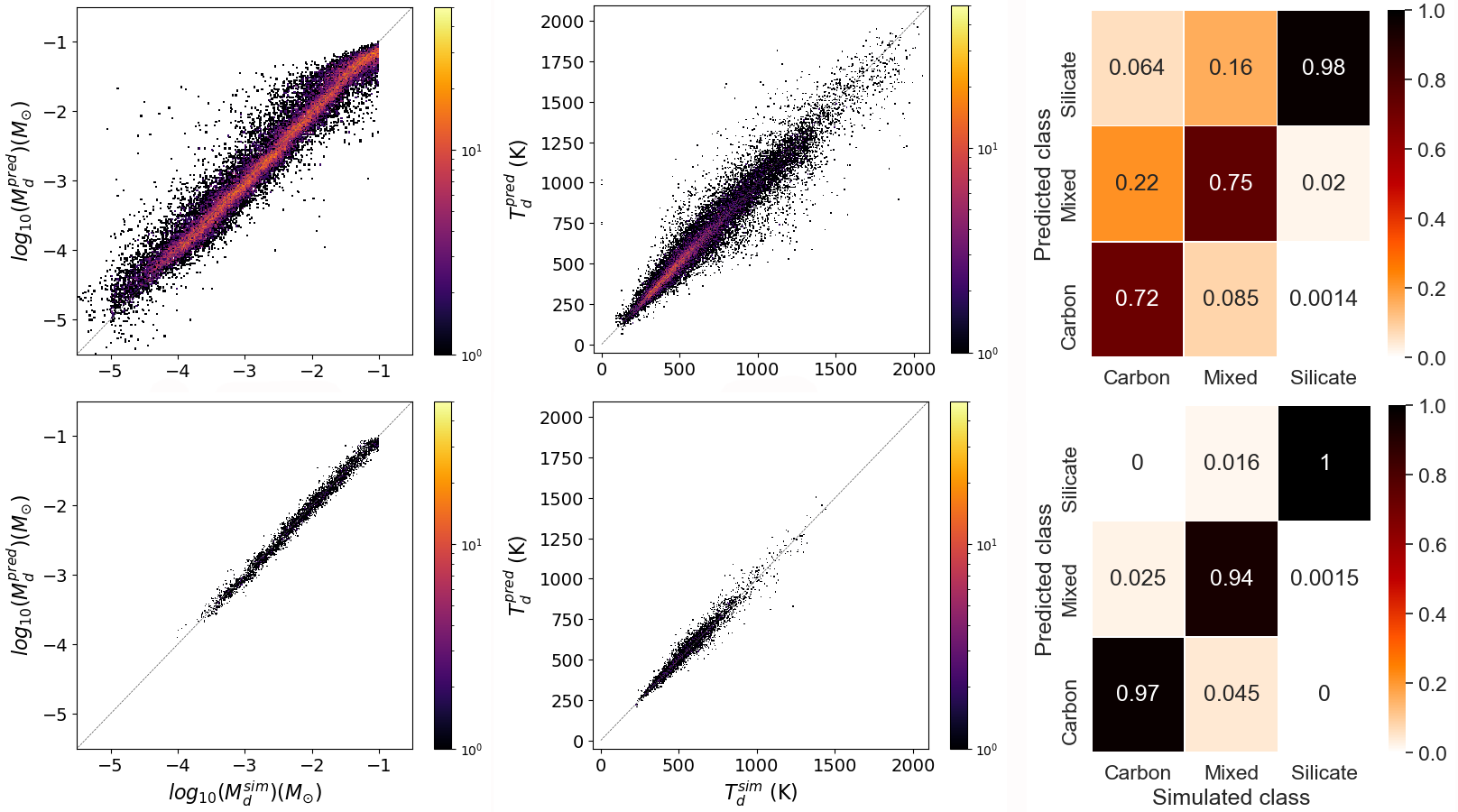}\\
    
    \caption{Performance of the neural network with the preferred subset of JWST filters, for S3. The estimations are shown for the reliable predictions of the test data set with S/N=3 ({\it top panel}), and the test data set with S/N=20 ({\it bottom panel}). The definition of the columns, the variables, the dashed lines and the colour bars are the same as in Figure~\ref{fig:1st_opt_test}.}
    \label{fig:3rd_opt_test}
\end{figure*}

In Figures~\ref{fig:1st_opt_test},~\ref{fig:2nd_opt_test}, and~\ref{fig:3rd_opt_test}, the performance of the neural network is shown for case-1 and case-2 for all scenarios.
Overall, the performance of the neural network for case-2 is better than for case-1.
As illustrated in the top panels of Figures~\ref{fig:1st_opt_test} and \ref{fig:2nd_opt_test}, the dispersion of the predictions around the diagonal line that represents predicted values equal to simulated values, increases from S1 to S2 for both target values \mdust\ and \tdust. Moreover, as summarised in Table~\ref{tab:clacc} the classification accuracy decreases for all \sdust\ from S1 to S3 in case-1.

As shown in Figure~\ref{fig:1st_opt_test}, for S1 the reliable predictions for \mdust\ and \tdust\ range between about $6 \times 10^{-5}-10^{-1}$ \Msun\ and 100--1\,400 K, respectively. However, Figure~\ref{fig:2nd_opt_test} shows that in S2, the reliable predictions only range between about $10^{-4}- 5 \times 10^{-2}$ \Msun, and 250--1\,200 K for \mdust\ and \tdust, respectively. Figure~\ref{fig:3rd_opt_test} shows that in S3, the reliable predictions for \mdust\ are within $5 \times 10^{-4}-10^{-1}$ \Msun\ and 250--1\,000 K for \tdust. This means that the dust mass and temperature range of the reliable predictions for all cases shrinks from S1 to S2 to S3, and thus with the increased complexity of the scenarios.

\subsection{Feature selection}
\label{ss:result_fs}
\begin{figure}
    \centering
    \includegraphics[width=0.5\textwidth]{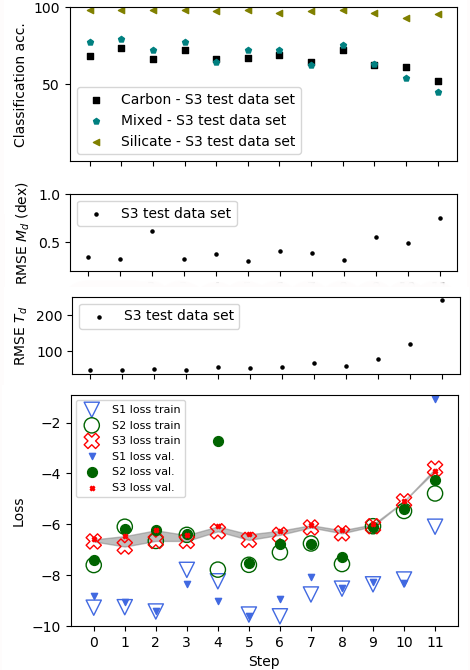}
    
    \caption{Performance of our neural network with each filter set that are obtained at each step of the feature selection process. {\it Bottom Panel}: Loss values that are achieved by the training and validation data sets at the end of each training process of the neural network in S1 (downward triangles), S2 (circles), and S3 (X symbols). The empty symbols mark the training loss for each step. The filled symbols mark the validation loss for each step. The grey shaded region represents the area between the training and validation loss in S3. The single panels show the RMSE of \tdust\ (K), and \mdust\ (\Msun), and the classification accuracy (\%) for predicting the dust species for the test data sets in S3, from bottom to top. The classification accuracy for predicting carbon and silicate dust species, and a mixture of them are shown with circles, triangles, and dashes respectively. }

    \label{fig:loss_all_shap}
\end{figure}

%all the losses
Figure~\ref{fig:loss_all_shap} presents the performance of the neural network with the subsets of the JWST filters that are selected in each step of the feature selection process. The bottom panel compares the training losses obtained for the last epoch at all feature selection steps for all three scenarios. The validation losses for S3 are also included in Figure~\ref{fig:loss_all_shap}. It is evident that for S3, the validation loss closely follows that of the training loss. We find the same for the other two scenarios, although the loss values vary more drastically from step to step. The absolute local minimum of both the validation and training loss appears to be reached in step zero of the feature selection process for S3, while for both S1 and S2 the absolute local minimum is reached in step five. However, we find for S3 that both the training and validation loss slowly increase from step zero to eight by about 5\%. 
 
The first three panels of Figure~\ref{fig:loss_all_shap} show the performance evaluation of the neural network for the test data sets. It is evident that the RMSE for \tdust\ varies only minimally around a mean value of about 58 $\pm$ 9 K, after which it increases to about 240 K in the last step. The RMSE of \mdust\ behaves similarly constant over the first eight steps except for step two, and increases from step eight to eleven by about 0.45 dex. The classification accuracy for carbon dust and the mixed composition also only changes minimally over the first eight steps, but appears to decrease from step eight to eleven from about 70\% to about 50\%. For silicate dust the classification accuracy remains nearly 100\% over all steps.

\begin{table}

\centering
\caption{Preferred and minimum subsets of JWST filters obtained from the feature selection process and used to estimate \mdust, \tdust, and \sdust. This is shown for all three scenarios. Columns termed `Pref.' and `Min.' stand for preferred and minimum subset of JWST filters, respectively. }
\begin{tabular}{p{1.8cm}|p{1.44cm}|p{1.44cm}|p{1.44cm}}
\hline \hline
JWST filters & S1 & S2 & S3\\
\end{tabular}
\begin{tabular}{p{1.8cm}|p{0.5cm}|p{0.5cm}|p{0.5cm}|p{0.5cm}|p{0.5cm}|p{0.5cm}}
 & Pref. & Min. & Pref. & Min. & Pref. & Min. \\
\end{tabular}
\begin{tabular}{p{1.8cm}|p{0.5cm}|p{0.5cm}|p{0.5cm}|p{0.5cm}|p{0.5cm}|p{0.5cm}}
\hline
$F2550W$ & $\bullet$ & & & & &
\\
$F2100W$ & $\bullet$ & $\bullet$ & $\bullet$ & & &
\\
$F1800W$& $\bullet$ & &$\bullet$ &$\bullet$ & $\bullet$ &$\bullet$
\\
$F1500W$& $\bullet$ & $\bullet$&$\bullet$ &$\bullet$ &$\bullet$ &$\bullet$
\\
$F1280W$& $\bullet$ & $\bullet$&$\bullet$ &$\bullet$ & $\bullet$ &
\\
$F1130W$& $\bullet$ & $\bullet$& & &$\bullet$ &$\bullet$
\\
$F1000W$& $\bullet$ & &$\bullet$ &$\bullet$ &$\bullet$ &$\bullet$
\\
$F770W$& $\bullet$ & $\bullet$&$\bullet$ &$\bullet$ &$\bullet$ &$\bullet$
\\
$F560W$& $\bullet$ &$\bullet$ &$\bullet$ &$\bullet$ &$\bullet$ &$\bullet$
\\
$F480M$&  & & & &$\bullet$ &$\bullet$
\\
$F470N$& & & & & &
\\
$F466N$& $\bullet$ & & & & &
\\
$F460M$& $\bullet$ & $\bullet$ & & & $\bullet$&
\\
$F444W$& $\bullet$ & & & & &
\\
$F430M$& $\bullet$ & & & & &
\\
$F410M$&  & &$\bullet$ &$\bullet$ & &
\\
$F405N$& & & $\bullet$& & &
\\
$F360M$& $\bullet$ & &$\bullet$ &$\bullet$ & &
\\
$F356W$&  & & $\bullet$ &$\bullet$ &$\bullet$ &$\bullet$
\\
$F335M$& $\bullet$ &  & & & &
\\
$F323N$&  & & & & &
\\
$F322W2$& &  & $\bullet$ & & &
\\
$F300M$& $\bullet$ &  & $\bullet$ &$\bullet$ & &
\\
$F277W$& & & $\bullet$ &$\bullet$ & &
\\
$F250M$&  & & $\bullet$ & & &
\\
$F212N$& & & & & &
\\
$F210M$& $\bullet$ & & & & &
\\
$F200W$&  & &$\bullet$ & & &
\\
$F187N$&  & & $\bullet$& & &
\\
$F182M$&  & &$\bullet$ & & &
\\
$F164N$& $\bullet$ & & & & &
\\
$F162M$&  & & & & &
\\
$F150W$& $\bullet$ & &$\bullet$ &$\bullet$ & &
\\
$F140M$& $\bullet$ & &$\bullet$ &$\bullet$ & $\bullet$ & $\bullet$
\\
$F115W$& $\bullet$ & &$\bullet$ & & $\bullet$ &
\\
$F090W$&  & & & & &
\\
$F070W$& $\bullet$ & &$\bullet$ & & $\bullet$&$\bullet$
\\
\hline
\end{tabular}
\\
\label{tab:opt-min-filters}
\end{table}

%%%%%%%%3rd scenario

%%%%%%%%%losses

%%%feature importance-3rd

\begin{figure}
    \centering
    \includegraphics[width=0.45\textwidth]{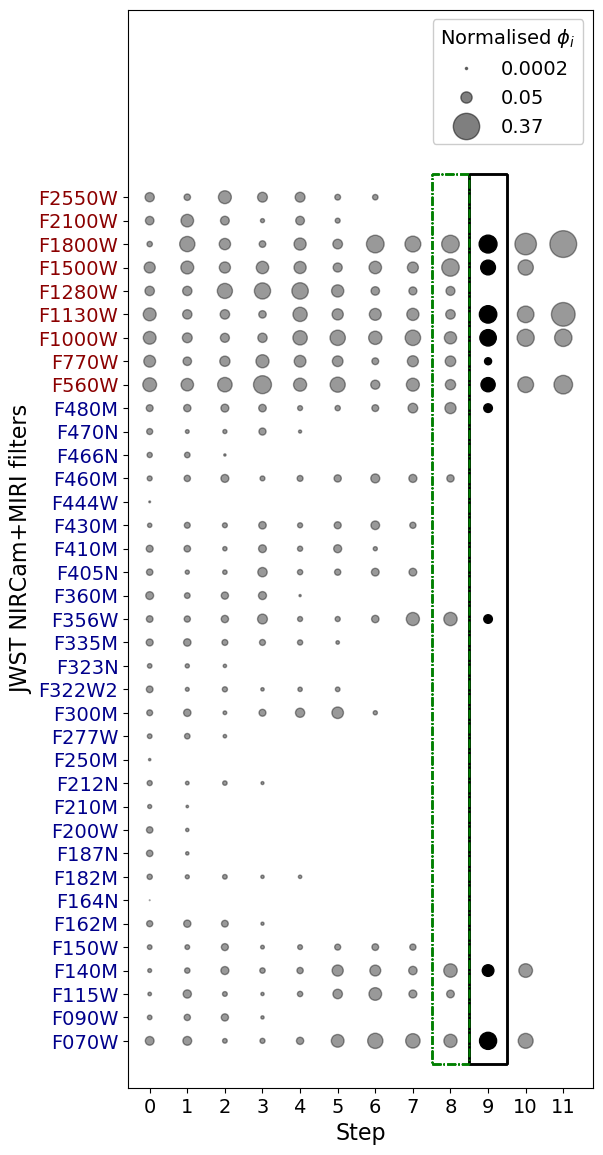}
    \caption{Importance of JWST filters for estimating the amount, temperature and the dust species, in S3. The normalised feature importance ($\phi_{i}$) of each NIRCam (blue) and MIRI (red) filters in each step of the feature selection process is shown by the size of the filled circles that are scaled to three values in the legend. The preferred and minimum subsets of JWST filters are highlighted with boxes using a dash-dotted and a solid line, respectively.}
    \label{fig:bands_its_s3}
\end{figure}

%%%%%%%%
%%%%%%%different S/N results
\begin{figure*}
    \label{fig:3rd_opt_test_s/n}
    \includegraphics[width=1\textwidth]{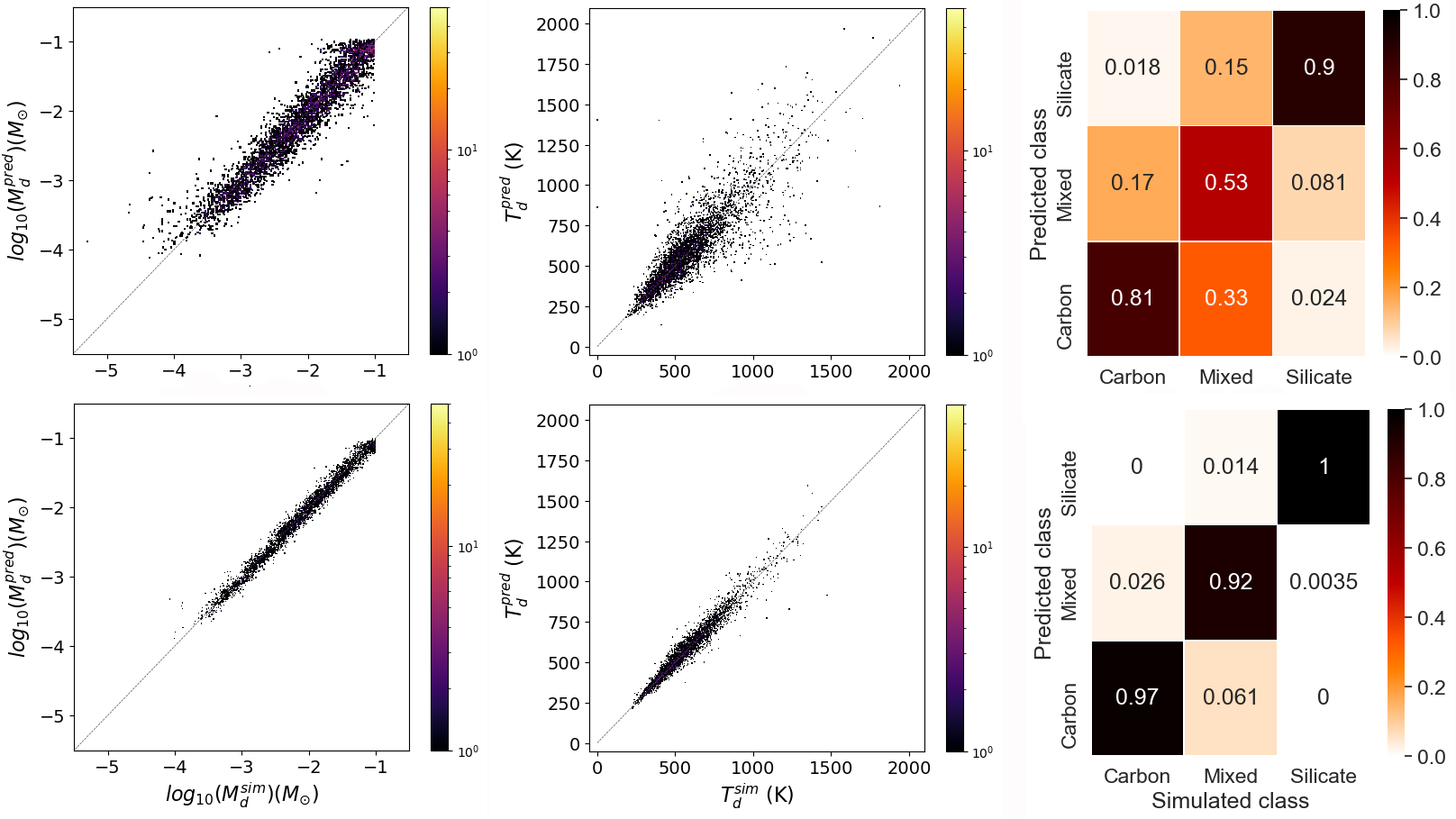}
    \caption{Performance of the neural network with the preferred subset of JWST filters, for S3.  The definition of the panels, the variables, the dashed lines and the colour bars are the same as in Figure~\ref{fig:1st_opt_test}.}

\end{figure*}

%%%-----tables for test data sets with different S/N

\begin{table*}
\caption{Comparison of neural network performance for estimating \mdust\ and \tdust\ in scenario 3 for case-2 with the same definition in Table~\ref{tab:mdtd}. Two test cases with test data sets with S/N=3 and S/N=10, respectively, are evaluated. In both test cases the training data set is similar to case-2. The evaluation metrics are applied on the reliable predictions from each test case.}
\centering
\label{tab:mdtd_s/n}
\begin{tabular}{ p{1.5cm}|p{1.5cm} p{2cm} p{2.5cm}| p{1.5cm} p{2cm} p{2.5cm}}          % centered columns
\hline\hline                        
 &   &\mdust~(\Msun) & & &\tdust\ (K) &

\end{tabular}
\\
\begin{tabular}{ p{1.5cm}| p{1.5cm} p{2cm} p{2.5cm}| p{1.5cm} p{2cm} p{2.5cm}} 
% \hline
S/N & bias (dex) & RMSE (dex) & 3$\sigma$ outliers (\%) & bias & RMSE & 3$\sigma$ outliers (\%)
\\
\hline

%  & S1\tablefootmark{d} &  &  &   &  &   & \\
 3 & 0.0047 & 0.4281 & 2.29 & 7.43 & 88.82 & 1.94\\
 20 & -0.0387 & 0.1237 & 15.02 & 5.10 & 32.08 & 1.36\\
   
\hline                                            
\end{tabular}
%\\
\end{table*}

\begin{table*}
\caption{Comparison of neural network performance for classifying dust species, and the fraction of predictions of the test data set that have \rsk\ to all the predictions from the test data set (\fracrsk), in scenario 3 for case-2 as defined in Table~\ref{tab:mdtd}. Two test cases with test data sets with S/N=3 and S/N=10, respectively, are evaluated.}
\centering
\label{tab:clacc_s/n}
\begin{tabular}{p{1.5cm} |p{0.9cm} p{5cm} p{0.1cm} |p{2cm}}
\hline \hline
&  &\clacc\ rate (\%) &&\\
\end{tabular}
\\
\begin{tabular}{ p{1.5cm} |p{2cm} p{2cm} p{2cm} |p{2cm}}
S/N & Carbon dust & Mixed dust  & Silicate dust &\fracrsk (\%)\\
\end{tabular}
\\
\begin{tabular}{ p{1.5cm} |p{2cm} p{2cm} p{2cm} |p{2cm}}
\hline

3 & 81 & 53 & 90 & 12 \\
20 & 97 & 92 & 100 & 12\\
\hline

\end{tabular}
\tablefoot{
{With the subset of JWST filters that are selected via the feature selection procedure as follows:}\\
\tablefoottext{c}{: See the definitions in Table~\ref{tab:mdtd}.}}
\end{table*}

%%%%%%%%%---------lowR-highR

\begin{figure}
    \centering
    \includegraphics[width=0.5\textwidth]{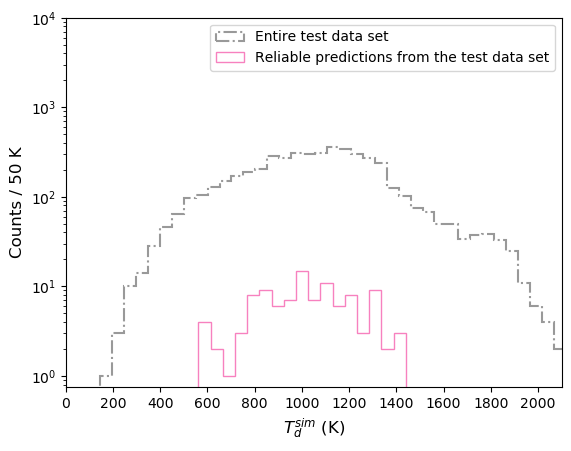}\\
    \includegraphics[width=0.5\textwidth]{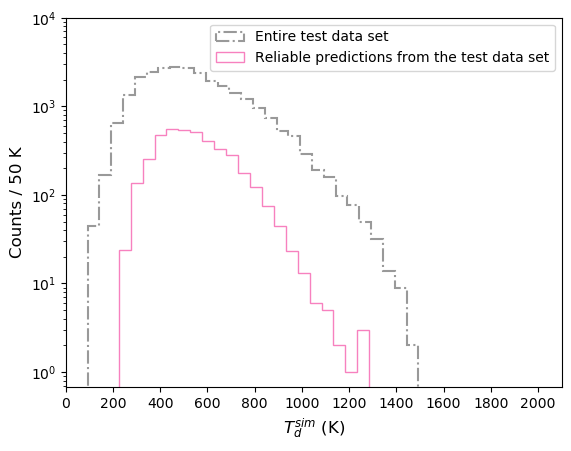}
    \caption{Distribution of SN model SEDs in \tdust\ in 50\,K bins. The dash-dotted line represents the distribution of the entire test data set and the solid line represents the distribution of the reliable predictions from the test data set.
    {\it Top panel}: \smodelss\ with $R_{in} \lesssim 5 \times 10^{16}$ cm. {\it Bottom panel}: \smodelss\ with $R_{in} \gtrsim 5 \times 10^{16}$ cm. }
    \label{fig:R}
\end{figure}

%------------------------------------------
\section{Discussion}
\label{s:discussion}
The performance evaluation of our trained neural network, which is designed to predict dust properties such as \mdust, \tdust\ and different \sdust, demonstrates that neural networks can be a powerful tool, if a sufficiently large data set is at hand. 
One advantage of using such a method is that it is possible to obtain a good estimate on the prediction uncertainties for each dust property under consideration. For other common methods, such as fitting a simple modified black body function or combination of thereof, uncertainties of the fitted dust mass or dust temperature are often not obtained \citep[e.g.][and references therein]{2011A&ARv..19...43G}. Furthermore, due to the fact that for such fitting methods assumptions about the dust composition need to be made a priori to fitting, the parameter range can be large and often not explored in all detail. The reasons for this may include insufficient data quality, but also time and computational limitations. These issues also apply to more sophisticated dust models such as \mocassin, when used to fit observational data to obtain the amount and temperature of dust in and around SNe  \citep[see e.g.][]{2015MNRAS.446.2089W}.

\subsection{Limitations of the model dataset}
\label{ss:limmoddatasets}

For the purpose of running a large number of models in a reasonable amount of time, we made some simplifying restrictions to the parameter space that our models cover. Some of these simplifications may have a significant effect on the predicted dust quantities from the SEDs. Our models used a single grain size only, selected from a uniform distribution in log-space. In the interstellar medium, the grain size distribution may be approximated by a \citet[][hereafter MRN]{1977ApJ...217..425M} distribution, in which the number density of grains of radius $a$ is proportional to $a^{-3.5}$. As this power-law distribution arises from collision and fragmentation processes over a long timescale, it is unlikely to be applicable to the dust grains found in and around CCSNe. A single grain size may be a more reasonable approximation. Observational studies tend to find evidence for large grains (e.g. \citealt{2014Natur.511..326G, 2015MNRAS.446.2089W,2015ApJ...801..141O, 2018A&A...611A..67B}). If a population of grains grows by accretion, then according to the standard grain growth equation, the increase in radius with time does not depend on the initial radius of the grain. A size distribution will therefore become narrower as accretion proceeds, unless fragmentation is also taking place.

\begin{figure}
    \centering
    \includegraphics[width=0.5\textwidth]{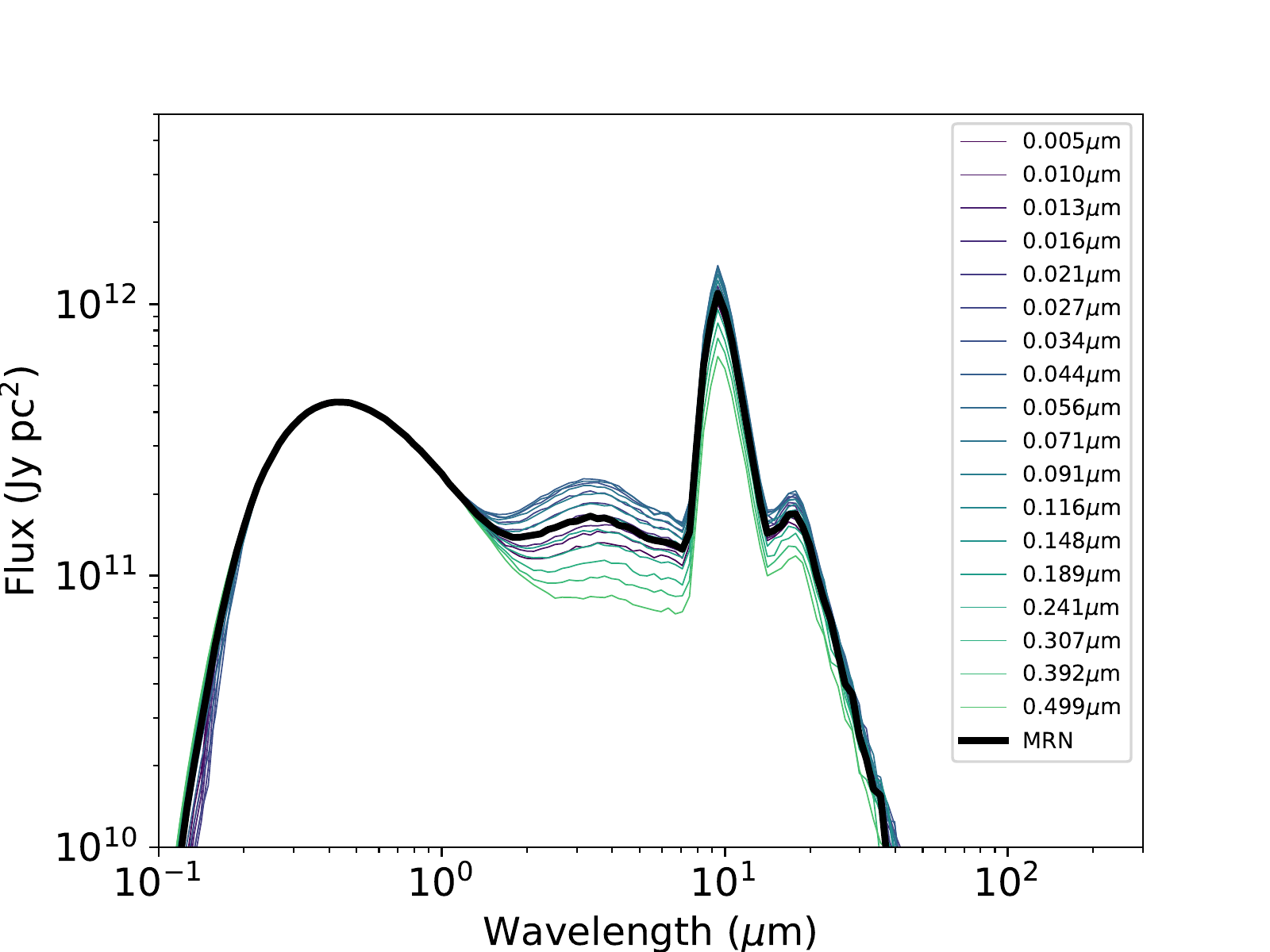}
    \caption{Comparison of predicted SEDs for single grain sizes between 0.005 and 0.5 $\mu$m. The colour coding is described in the legend of the figure. The SED for a grain size distribution with 0.005 $\mu$m $<a<$ 0.5 $\mu$m, n(a) $\propto$ a$^{-3.5}$ is shown as black solid line.}
    \label{fig:grainsizedistributiontest}
\end{figure}

In Figure~\ref{fig:grainsizedistributiontest}, to illustrate the effect of using a single grain size as opposed to a distribution, we show the predicted SEDs for 20 models characterised by a single grain size, evenly spaced logarithmically between 0.005 $\mu$m and 0.5 $\mu$m, together with the predicted SED for an MRN dust distribution. It can be seen that the SED for the full grain size distribution is almost identical to the SED for a single grain size of 0.15 $\mu$m.

The calculation of a spectral energy distribution from thermal dust emission fundamentally depends on the choice of optical constants. Different literature sets of optical constants may differ significantly from each other, and the dust actually present in and around a SN may not be well represented by the materials from which optical constants have been determined. The choice of optical constants thus introduces a systematic uncertainty into the dust mass and temperature estimates.

\begin{figure}
    \centering
    \includegraphics[width=0.5\textwidth]{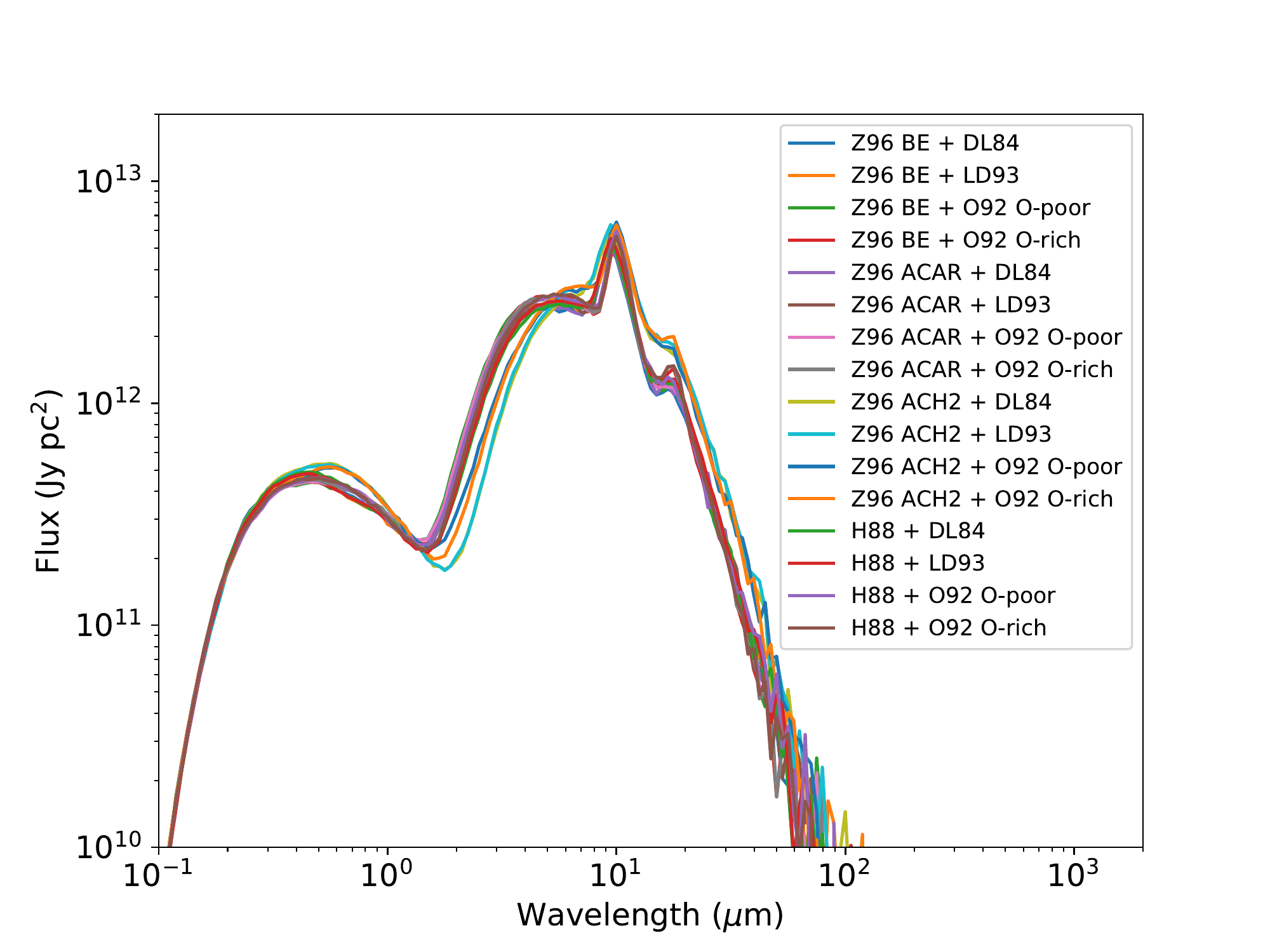}
    \caption{Effect of choice of optical constants on the predicted SEDs for a model with mixed composition. 16 SEDs are plotted, using all combinations of four sets of carbon and four sets of silicate optical constants as defined in Section~\ref{ss:limmoddatasets}.}
    \label{fig:opticalconstantstest}
\end{figure}

In a future work, we plan to investigate this more thoroughly, by using the neural network to classify SEDs calculated using different optical constants to those on which the network was trained. However, in this work, we used only two species of dust, and only one set of optical constants for each species. Dust in SNe is widely assumed to be either carbonaceous, silicaceous, or a mixture, and our SEDs are calculated using widely-used optical constants for these species. However, different choices of optical constants can yield significantly different SEDs. To illustrate this, we show in Figure~\ref{fig:opticalconstantstest} the variations in predicted SEDs for one example model. The model has a 50:50 silicate:carbon grain mixture, and Figure~\ref{fig:opticalconstantstest} shows the predicted SEDs for a single grain size of 0.1$\mu$m, using all possible combinations of optical constants from four sets of carbon data (\cite[][thereafter H88]{1988ioch.rept.....H}, and the ACAR, ACH2 and BE\footnote{These designations refer to amorphous carbon grains produced by arc discharge between amorphous carbon electrodes in an argon atmosphere (ACAR), arc discharge in a hydrogen atmosphere (ACH2), and burning of benzene in air (BE)} samples from \citealt[][thereafter Z96]{1996MNRAS.282.1321Z}) and four sets of silicate constants (\citealt[][thereafter DL84 and LD93, respectively]{1984ApJ...285...89D, 1993ApJ...402..441L}, and oxygen-deficient and oxygen-rich constants from \citealt[][thereafter O92]{1992A&A...261..567O}).

It is clear from Figure~\ref{fig:opticalconstantstest} that different choices of optical constants can result in significant differences in some wavelength regions of some SEDs. Particularly affected appears to be the 1--10 $\mu$m regions. However, the differences are largest for relatively small grains and are negligible for grains as large as 5$\mu$m. As mentioned, many observational studies of dust in young and old SNRs have found evidence for generally large grains, thus tending to reduce the uncertainty due to the choice of optical constants. Additional comparisons for more grain sizes and for pure carbon and pure silicate compositions are given in Appendix~\ref{app:opticalconstants}.

\subsection{Performance evaluation}
Our performance evaluation demonstrates that for all scenarios and cases (see Table \ref{tab:mdtd}), the obtained prediction error, RMSE, for \mdust\ is smaller than $\sim$ 0.55 (dex), and is smaller than $\sim$ 78 K for \tdust. These RMSE are obtained for case-3 and S3 and are the maximum RMSE values out of all scenarios and cases. 
This is because in case-3, the evaluation metrics are applied only onto the test data sets and minimum subsets of JWST filters of each scenario S1, S2 and S3. 
Moreover, for the evaluation of case-3 the test data sets are used without prior \rsk\ cut and hence, contain predictions with larger uncertainties. Additionally, S3 is the most complex scenario of all scenarios. However, compared to other works in the literature with inferred amounts and temperatures of dust from observed SNe,
we find that even the worst performance here in this work constitutes a very good performance.
For example, we can compare to other works that estimate the amount of dust, with {\it Spitzer Space Telescope} observations up to about 25 $\mu$m for SNe such as SN 2004et \citep{2009ApJ...704..306K} and SN 1987A  \citep{2007MNRAS.375..753E}.
For SN 2004et, the estimated range for dust mass and dust temperature at 300, 464 and 795 days after the explosion, are about 0.37 dex and 500 K, 0.26 dex and 250 K, and 0.38 dex and 80 K. For SN 1987A, the amount of carbon dust at day 615 has been estimated with an uncertainty of 0.81 dex.

We now turn to the performance evaluations of the most reliable predictions, which are drawn from case-1 and case-3 data sets that have \rsk, evaluated as case-2 and case-4. Comparing the RMSE for \mdust\ and \tdust\ between case-1 and case-2 (cases with the data sets that contain the preferred subsets of JWST filters) across the three scenarios, S1, S2 and S3, shows that the prediction errors are reduced by up to a factor of about 2--3 in case-2 where the predictions that do not have \rsk\ are excluded. 
Since this is an expected, but not guaranteed, consequence of including only the predictions that have \rsk, which removes `bad' predictions that do not fulfil the criterion to have \rsk, the same is expected for the cases with the data sets that contain the minimum subsets of filters (case-3 and case-4). Our evaluations show that the effect of excluding the unreliable predictions for \mdust\ estimations is even stronger than that for \tdust, meaning that the RMSE (in dex) of the dust mass is smaller by about a factor of 4--5 in case-2 and case-4, compare to case-1 and case-3, while for \tdust\ the decrease is only about factor of 2. The classification accuracy of classifying the different dust species shows the same behaviour, which is higher for nearly all species and scenarios for case-2 and case-4 than for case-1 and case-3. 
Particularly for silicate dust, the classification accuracy is close to or at 100\%. However, in case-4 and S3, there is a bias in predicting the mixed dust species towards the carbon dust species.
The evaluation method using the \rsk\ definition demonstrates that the dust mass and temperature predictions that have been under the scrutiny of the \rsk\ criterion can truly be considered as reliable predictions.

On the other hand, as shown by \fracrsk\ in Table~\ref{tab:clacc}, the number of predictions that satisfy the \rsk\ criteria in case-2 and case-4 is smaller than the predictions using the entire data set as in case-1 and case-3. Since there are $3\sigma$ outliers (as defined in Section~\ref{s:evaluation}) also for case-2 and case-4, the fractions of the best reliable predictions for S1, S2 and S3 in case-2 and case-4 are smaller than \fracrsk. For instance, in case-2 for S1 the fraction of the best reliable predictions is still about 59\% while in case-4 and S3 it shrinks to only about 5.8\%. 

Comparing the number of $3\sigma$ outliers between \mdust\ and \tdust, we find that for nearly all setups of cases and scenarios, the \mdust\ evaluations result in a larger number of $3\sigma$ outliers than the \tdust\ evaluations. This is because the dispersion of \mdust\ residuals is larger than \tdust\ residuals.  

\subsection{Filter selection}
\label{ss:filter_selection}

Since observing SNe with all the JWST filters at the same time is practically not feasible, we are interested in finding the smallest set of filters with which an acceptable performance can be achieved. To do so, we utilised a feature selection process as described in Section~\ref{ss:featselect}. From this we obtain two sets of filters for each scenario, one preferred set of filters and one minimum set of filters. The preferred filter set is chosen based on the absolute minimum reached by both the training and the validation loss while the minimum filter set is chosen based on criterion that the fraction of the number of reliable predictions to the total number of predictions is larger than 5\%.
It turns out that for S1 and S2 the preferred filter set is reached early in the filter selection process, step five, and thus still contains a large number of filters (22 filters). The minimum filter set is obtained in steps eight or ten, and thus contain fewer, between seven to thirteen, filters. Looking at the performance evaluation from the two filter sets, for example case-1 and case-3 or case-2 and case-4 in Table~\ref{tab:mdtd}, then while as expected, the performance of the neural network is overall better with the preferred set of filters. The performance with the minimum filter set is only minimally decreased. Hence, as demonstrated in Table~\ref{tab:mdtd}, accurate predictions of \tdust\ and \mdust\ can be achieved with the minimum set of filters.    

For scenario S3, the preferred and the minimum subset of filters are chosen from step eight and nine and are thus very close to each other. It is important to note that in this case the preferred set is chosen to be at step eight instead of step zero, where the loss reached the absolute minimum. However, we do not consider step zero as `preferred'. Since both the training and validation loss remains rather stable until step eight as pointed out in Section~\ref{ss:result_fs}, step eight can be considered as preferred. 

As illustrated in Figure~\ref{fig:loss_all_shap}, in S3 compared to S2 there are insignificant changes of loss values in each step of the feature selection process up to step nine.
This stability of the performance of the neural network in S3, regardless of the number of filters that are used as the input features can be due to the training of the neural network with additional noise. This is because the training of a neural network with additional noise can be equivalent to a regularisation \citep{6796505}, which helps the neural network to react less to the variation of input features. Therefore, in S3 compared to S2, the training and validation losses that are achieved by the neural network with smaller sets of filters than the entire filter set, do not significantly change in each step of the feature selection process up to step nine.

Figure~\ref{fig:bands_its_s3} visualises the resulting Shapley values obtained for each step in scenario S3. Figures~\ref{fig:bands_its_s1}, and ~\ref{fig:bands_its_s2} show the same for S1 and S2, respectively. It is interesting to note that for all three scenarios, none of the narrow-band JWST filters are amongst the minimum subsets of the JWST filters. However, for S1 and S2 two such narrow-band filters are included in the preferred filter set albeit with small Shapley values and hence, marginal importance.  

This implies that real observations of SNe with such JWST narrow-band filters would have the least impact on estimating dust properties with our neural network.
As shown in Figure~\ref{fig:bands_its_s3}, and expected, the MIRI filters that cover the longer wavelength region are crucially important to estimate the dust properties while the shorter wavelength NIRCam filters seem not to play a significant role. 

One of the most pressing questions of course is, if it is technically feasible to construct an observing run with the minimum subset of filters.
% : `Is it technically feasible to construct an observing run with the minimum subset of filters?'  
The NIRCam instrument uses a dichroic to split the incoming radiation into two wavelength ranges, $\lambda$ < 2.5 $\mu$m and $\lambda$ > 2.5 $\mu$m, known as short and long wavelength channels \citep{10.1117/12.552281}.
% (Reference if possible). 
This setup allows to simultaneously obtain two images with two different filters, each from one of the channels. Since, in the minimum subset for S3, two selected NIRCam filters, $F070W$ and $F140M$, are in the short wavelength channel of NIRCam and two, $F356W$ and $F480M$, are in the long wavelength channel, two separate runs are required to observe a SN with all four NIRCam filters.
For MIRI, observations can only be conducted with one filter at a time. The entire observing time needed for all selected MIRI filters of the preferred subset may in the end depend on the brightness of the SN, the either desired or best possible signal-to-noise ratio or the phase of the SN.

\subsection{Additional testing of the performance of the neural network}

Our simulated data set is simplified by various assumptions such as a uniform S/N=10. In reality, the achieved S/N ratio depends on different aspects, such as the brightness of the object in a given filter band, the distance to the object or the exposure time and integration setup. The JWST Exposure time calculator is an ideal tool to adjust all these aspects. It is obvious that for bright sources a high S/N ratio even with a short exposure is possible to achieve, while for faint sources, long exposure times may be necessary to reach just a minimum significance of S/N$\approx$3. While simulating more realistic S/N ratios for each filter band assuming different exposure times is possible, it is computationally expensive and hence, we decided to first test the neural network performance for a simple case, a uniform S/N=10, which represents neither particularly good nor bad data. 

However, to better understand the effect of better or worse data, we tested the performance of our neural network for scenario S3 and case-2 on two test cases with test data sets assuming S/N=20 and S/N=3, respectively. The results are summarised in Tables~\ref{tab:mdtd_s/n} and~\ref{tab:clacc_s/n} and presented in Figure~\ref{fig:3rd_opt_test_s/n}. For the test case representing higher quality data with a S/N=20, the RMSE of \mdust\ and \tdust\ are  $\sim$0.12 M$_{\odot}$ (dex) and $\sim$32\,K. The RMSE of \mdust\ and \tdust\ predictions for the other test case with a test data set with S/N=3 are $\sim$0.42 M$_{\odot}$ (dex) and $\sim$88\,K.   
As expected, the performance of the neural network has become worse for the test-case with a S/N=3 compared to the S/N=10 while for the test-case with a S/N=20 the performance remains similar. We note that since the neural network has been trained for a S/N=10, for the test-case of S/N=3 our predictions are somewhat over-confident while they are under-confident for S/N=20.

The final test of the usability of our neural network, which has been trained on a simulated data set exploring a wide, but not exhaustive range of parameters, is to use true observational data. Hence, we used the spectrophotometric observations of SN~1987A taken with the Kuiper Airborne Observatory at 615, 632 and 638 days past explosion (referred to as 615 day epoch) \citep{1993ApJS...88..477W, 1989Natur.340..697M}, as this epoch shows a clear signature of dust formation in the ejecta. The data cover a wavelength range of 0.33 -- 29.5 $\mu$m. Furthermore, \citet{2015MNRAS.446.2089W} has also fit \mocassin\ models to the same data and their best fit results in about 1 $\times$ 10$^{-3}$ \Msun\ of dust for a clumpy model with a 85:15 carbon:silicate ratio and temperatures of 252 $\pm$ 29\,K for carbon, and 316 $\pm$ 31\,K for silicate dust. This is a larger dust mass than the best fitting models by \citep{2007MNRAS.375..753E} who obtained $\sim$ 2 $\times$ 10$^{-4}$ \Msun\ at similar temperature, while \citep{1993ApJS...88..477W} obtained about 3.1 $\times$ 10$^{-4}$ \Msun\ at about 400 K of graphite dust, assuming a smooth dust distribution. 

Here, we created a small test data set which consists of SN~1987A data at 615 days that were replicated 500 times, each assigned a different redshift which was chosen randomly from a limited redshift range (0.0006-0.004). This ensures that the data are within the saturation and detection limits. We applied a Gaussian smoothing operator, which enabled interpolation between the data gaps at 1.02 -- 1.48 $\mu$m and 12.67 -- 17.32 $\mu$m and convolved the data with the JWST bandpass filters. We used the trained neural network of scenario S3, first, including all JWST bandpass filters (scenario S3) and second, using only the preferred set of the filters (S3, case-2) to predict the dust mass and temperature as well as the dust grain composition (carbon, silicates or 50:50 mix).

\begin{figure*}
    \centering
    \includegraphics[width=0.9\textwidth]{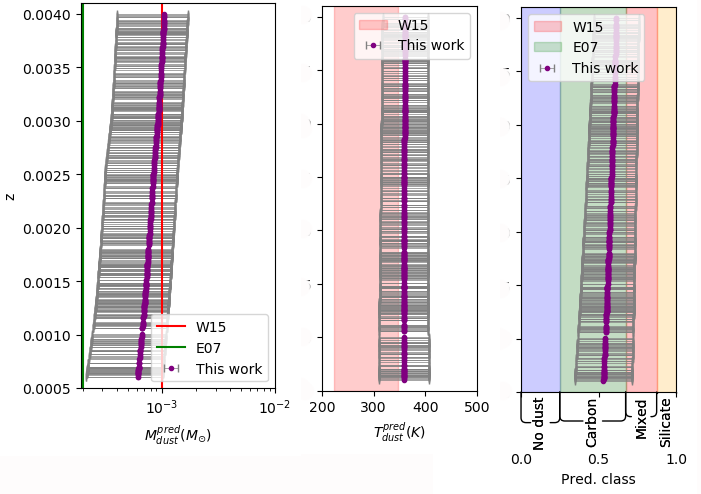}
    \caption{Estimated amount, temperature, and composition of the dust in SN 1987A at 615 days after explosion for the entire set of JWST filters. The purple dots along with black lines represent the predicted values and the predicted uncertainties by the trained neural network, respectively. The estimated values by \cite[][W15]{2015MNRAS.446.2089W} and \citep[][E07]{2007MNRAS.375..753E} are shown as red and green solid lines (left panel) and shaded areas (middle and right panels). The blue and yellow regions in the right panel highlight x-axis labels; No dust and Silicate.}
    \label{fig:all_filters_prediction}
\end{figure*}
\begin{figure*}
    \centering
    \includegraphics[width=0.9\textwidth]{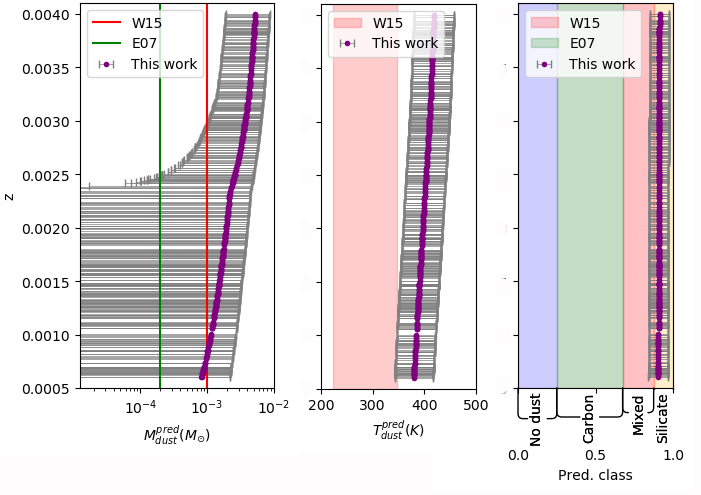}
    \caption{Estimated amount, temperature, and composition of the dust in SN 1987A at 615 days after explosion for the preferred set of JWST filters. The symbols, lines and shaded regions are defined as in Figure\ref{fig:all_filters_prediction}}
    \label{fig:preferred_filters_prediction}
\end{figure*}

The results are shown in Figures~\ref{fig:all_filters_prediction} and \ref{fig:preferred_filters_prediction}. There appears a trend with redshift for all predictions in all two cases. We find that with increasing redshift, the dust mass and temperature predictions increase and the predicted dust species is leaning more towards silicates. 
Using all JWST bandpass filters, we obtain dust masses that are predicted with 99.7\% confidence to range between a few times 10$^{-4}$ -- 10$^{-3}$ \Msun\ and temperatures to range between $\sim$ 280 -- 340\,K, in agreement with previous estimates in the literature. The estimated dust species is carbon or a mix of carbon and silicates. 
In the case of using only the preferred filter set (see Section~\ref{ss:filter_selection}), at very nearby distances, the results show that with a 99.7\% confidence the predicted dust mass is not larger than 2 -- 4 times 10$^{-3}$ \Msun\ while at z $\gtrsim$ 0.003 the predicted dust mass ranges from 10$^{-3}$ -- 10$^{-2}$ \Msun\ for a predicted dust species that can either be mixed or silicates. For all predictions, the temperature range overlaps with that from the first case using all JWST filter bands. 

This shows that our dust mass and temperature predictions for SN~1987A at 615 days are comparable to those in the literature and hence, our dust mass and temperature predictions are reasonable for SN~1987A-like SNe.
However we note that while the 
dust temperature predictions fulfil the \rsk\ criterion,
the dust mass predictions do not. Moreover, using the preferred JWST filter set, results in silicates as the dominant dust species, which disagrees with what is found in the literature. A possible reason for this may be ascribed to our simplified training data set and limited parameter range. Despite this, our method can be a promising tool to analyse  signatures of dust in and around SNe in their SEDs. In forthcoming work, we aim to use more detailed and realistic simulations to achieve more reliable predictions of the dust mass, temperature and possibly other dust properties.

\subsection{Implications for future observations}

Figure~\ref{fig:R} shows the histogram of the number of test data over \tdust, in S3, for the entire test data set (dashed line), and the sub-sample of test data with reliable estimated standard deviations (solid line). In the top panel, the distribution is shown for the \smodelss\ with R$_{\rm{in}} \lesssim 5 \times 10^{16}$ cm, while the bottom panel represents the \smodelss\ with R$_{\rm{in}} \gtrsim 5 \times 10^{16}$ cm. This cutoff represents an approximate division of models into those, which are closest to dust signatures from newly formed dust in young SNe ejecta, and those, which can be interpreted as signatures arising from
pre-existing circumstellar dust, flash-heated by a SN explosion. Pre-existing dust grains at radii less than about 5 $\times 10^{16}$ cm are likely to be evaporated by the SN explosion \citep{2014Natur.511..326G}, although some dust may survive. Meanwhile, SNe ejecta expanding with a mean velocity of $\sim 6\,000$ km/s would reach this radius after $\sim$1\,150 days. A SN following the bolometric evolution of SN~1987A \citep{2014ApJ...792...10S} would have a luminosity of $\sim$10\,000 L$_\odot$, the lowest considered in our models, at a similar epoch. Therefore, any dust estimated from model SEDs with dust located at distances $\gtrsim 5 \times 10^{16}$ cm (i.e. the top panel in Figure~\ref{fig:R}) may be interpreted as being
pre-existing dust. Dust estimates from model SEDs with dust located at distances $\lesssim 5 \times 10^{16}$ cm (i.e. the bottom panel in Figure~\ref{fig:R}) could be associated with 
newly formed dust at early epochs in young SNe. Such newly formed dust can either be located in the SN ejecta, or in the case of Type IIn SNe such as SN 2006jc \citep[e.g.][]{2008ApJ...686..467S} or SN~2010jl \citep[e.g.][]{2014Natur.511..326G, 2020ApJ...894..111B}, in the cool dense shell located at a distance of about $10^{16}$ cm and behind the forward shock which propagates through the dense circumstellar material that was shed off by the progenitor prior to the terminal explosion.
By comparing the covered areas in both panels, we find that our neural network may be better at estimating the dust mass and temperature of model SEDs which are more closer to a pre-existing dust scenario than a ejecta dust scenario. %
Whether or not this is due to our chosen simplifications and parameter space coverage of our simulations, can be tested in forthcoming, more realistic simulations.

In this work we used point source continuum sensitivity limits  \citep[][and described in Section~\ref{ss:jwstsenbri}]{2015PASP..127..686G, 10.1117/1.JATIS.3.3.035001} which are calculated assuming average zodiacal background levels. While this is a reasonable approach for young SNe that are in for example resolved nearby galaxies or maybe located at the outskirts of a galaxy or intergalactic medium, it may be problematic for older SNRs that are more extended, diffuse sources, as well as for SNe located in crowded regions. ISM back- and foreground contamination from unresolved stars in distant galaxies can give rise to a brighter background than assumed here, changing the sensitivity limits to cover lower magnitudes.  \citet[][]{10.1117/1.JATIS.3.3.035001} estimated that the sensitivity levels can worsen by up to a factor of $\sim$2 for NIRCam broad band filters in case of bright backgrounds. Contamination due to cold ISM dust with temperatures $\lesssim$ 30--50\,K could also affect the sensitivity limits. However, this is most prominent at longer wavelengths, $\gtrsim$ 100 $\mu$m, and thus may not significantly affect the sensitivity limits in JWST's wavelength range. Finally, the chosen observing strategy and possibilities for proper background subtractions may also shift the limits at which faint sources can still be detected. In forthcoming work we will test in more detail the impact of varying sensitivity limits on optimising the JWST filter selection to determine dust properties.

To use modern machine learning algorithms effectively, large data sets are essential. Presently ongoing wide-field surveys such as the Zwicky Transient Facility  \citep{2014htu..conf...27B}, Young Supernova Experiment \citep{2021ApJ...908..143J} or SkyMapper Southern Sky Survey \citep{2017PASA...34...30S} are discovering hundreds to thousands of SNe and other transients per year and are building up a wealth of optical photometric as well as spectroscopic data of various different types of CCSNe that will be further advanced in future surveys such as the Vera C. Rubin Observatory Legacy Survey of Space and Time \cite[][]{2019ApJ...873..111I}. While near- to mid-IR observations of CCSNe will likely boom with the launch of JWST and possibly other, future instruments on ground- and space-based telescopes, they are rare at present, and will most likely not reach the level required to train machine learning algorithms on observational near- to mid- to far-IR data.

The Open Supernova Catalog\footnote{\href{https://sne.space}{\color{blue}https://sne.space}} reported the discovery of 450 Type II, 102 Type Ib, and 60 Type Ic SNe in the year 2021, from which only a few have mid-IR data. Although this is a large number of observed CCSNe in just one year, collecting a data set of the size, wavelength range and degree of variation used in our study will also in future not be easily feasible. This `data-size' limitation is especially important for estimating the dust properties of the types of SNe we used here in this work, where we had simulate 293\,236 SN SEDs covering a wavelength range from 0.7 to 30 $\mu$m. Finally, as the dust properties and quantities cannot directly be measured from the observational data, thus are unknown, well advanced simulations with known dust properties are highly valuable. 
Therefore, applying a neural network that is well trained on a rich set of highly advanced simulated data exploring a large parameter space may be a promising way to determine dust quantities and properties of future observations. This work also allows testing in what detail quantities and properties of dust can be inferred from observational data. Furthermore future observational data, if included in the training of the neural network, can be used to validate the neural network and thus, will improve its performance and outcome.

%------------------------------------------
 \section{Conclusion}
 \label{s:conclusion}

In this work, we present a first test for using neural networks to estimate different quantities and properties of dust located in and around SNe including their predicted uncertainties.  
We aimed at predicting the temperature and amount of dust and to differentiate between three dust compositions.  
To do so, we simulated an extensive data set of 293\,236 \smodelss\ using the 3D photoionisation and dust radiative transfer code \mocassin\ \citep{2003MNRAS.340.1136E, 2005MNRAS.362.1038E}. We convolved the simulated data set with JWST MIRI and NIRCam bandpass filters. We considered the instrument's detection limits as well as estimated magnitude uncertainties to make the trained neural network suitable for predicting some of the properties and quantities of dust in SNe from future observations of this instrument. We defined three different scenarios to examine the feasibility and accuracy of inferring the dust properties by our neural network. In the first scenario, we assumed that all \smodelss\ have the same low redshift. In the second and third scenarios, we distributed all \smodelss\ within the redshift range of $0.0001-0.015$, in which at least seven JWST bandpass filters of all \smodelss\ are within the sensitivity and saturation limits that are calculated for a S/N of 10. Additionally, in the third scenario, we added random noise to the distributed \smodelss\ within a redshift range of $0.0001-0.015$. Thereafter, we selected the preferred and minimum subset of JWST filters from the feature selection process, which is based on the SHAP framework. We used these filter subsets to estimate the amount, temperature and dust species with our neural network.\\

From the outcome of our trained neural network in S3, which is the closest scenario to real observations, we find the minimum subset of JWST filters needed to estimate dust quantities and properties consists of NIRCam: $F070W$, $F140M$, $F356W$, $F480M$, and MIRI: $F560W$, $F770W$, $F1000W$, $F1130W$, $F1500W$, and $F1800W$ filters.
As presented in Table~\ref{tab:mdtd}, our neural network can well predict the dust quantities and properties for approximately 7\% of SN model SEDs from the entire test data set. This fraction has a RMSE of $\sim 0.12$ dex, and $\sim 38$ K for \mdust\ and \tdust. The classification accuracy is 95\%, 99\% and 57\% for carbon, silicate and a mix of carbon and silicate dust, respectively.
We find that the dust quantities and properties are best predicted by our neural network for \smodelss\ that approximately range in \tdust\ between 250--1000\,K, and \mdust\ between $5 \times 10^{-4} - 10^{-1}$ \Msun, and are dominated by astronomical silicates. %Given our simulated data set, these \smodelss\ can be associated with a pre-existing dust scenario.

%------------------------------------------
\begin{acknowledgements}
    We thank Dr. Doogesh Kodi Ramanah, Dr. Adriano Agnello and Dr. Mikako Matsuura for helpful discussions. We also like to thank the anonymous referee for insightful comments. This work is supported by a VILLUM FONDEN Investigator grant (project number 16599) and a VILLUM FONDEN Young Investor Grant (project number 25501). RW acknowledges support from European Research Council (ERC) Advanced Grant 694520 SNDUST. This work has made use of the Horizon Cluster hosted by Institut d’Astrophysique de Paris, and an HPC facility funded by a grant from VILLUM FONDEN (project number 16599)
\end{acknowledgements}

%------------------------------------------
\bibliographystyle{bibtex/aa}
\bibliography{bibtex/bibliography.bib}
%-----------------------------------------

%------------------------------------------

\begin{appendix}
\section{Sensitivity and saturation limits for NIRCam and MIRI}
\label{app:sens}
\begin{figure*}
    \centering
    \includegraphics[width=1\textwidth]{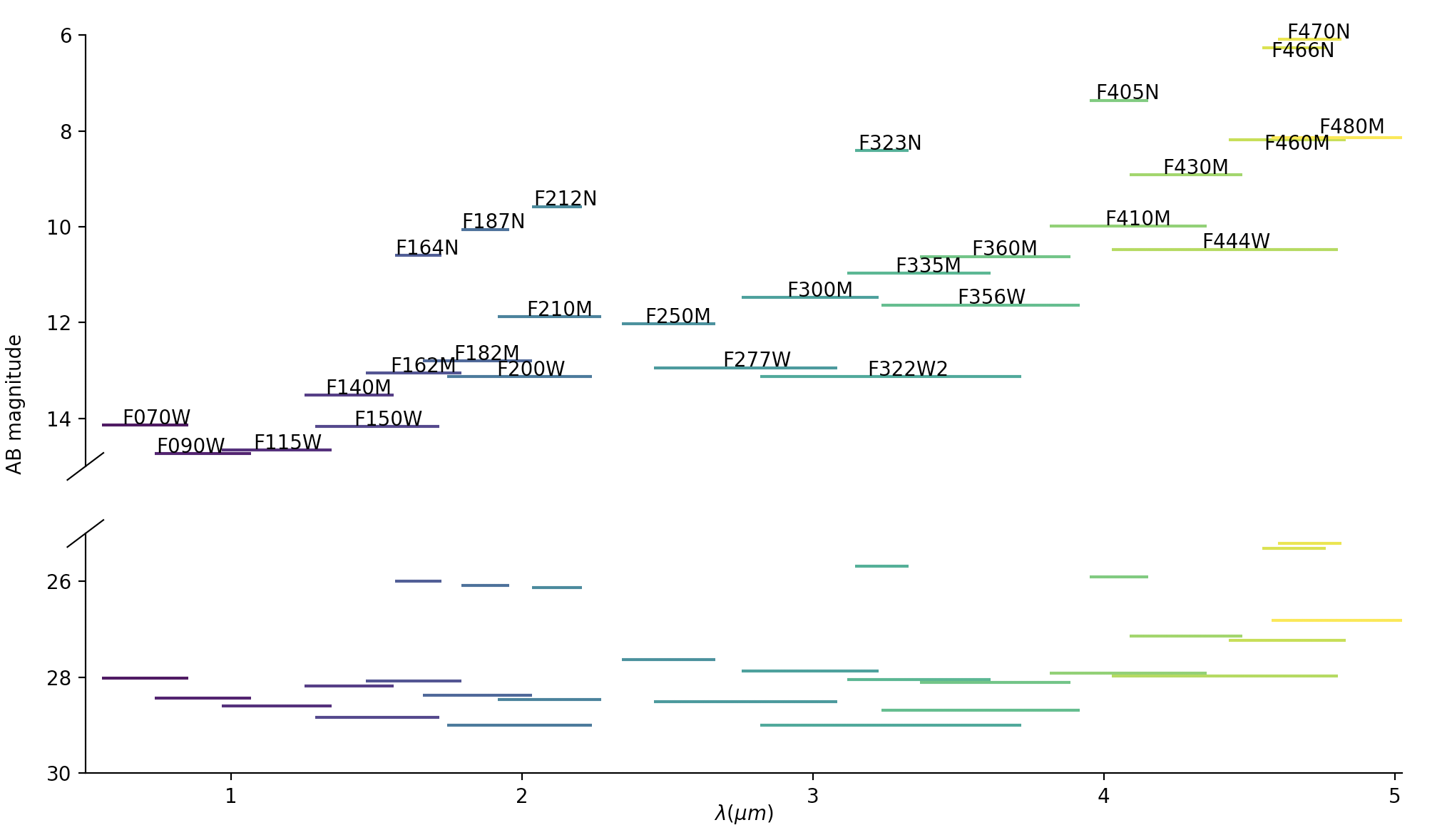}
    \caption{NIRCam saturation magnitudes in 10\,000 seconds exposure time, and point source sensitivity for 21.4 seconds exposure time. The sizes roughly represent the wavelength range of each filter. There is a break in y-axis (15--25 AB magnitude) to save a large blank space between the sensitivity and saturation limits.}
    \label{fig:sen_bri_NIRCam}
\end{figure*}

\begin{figure*}
    \centering
    \includegraphics[width=1\textwidth]{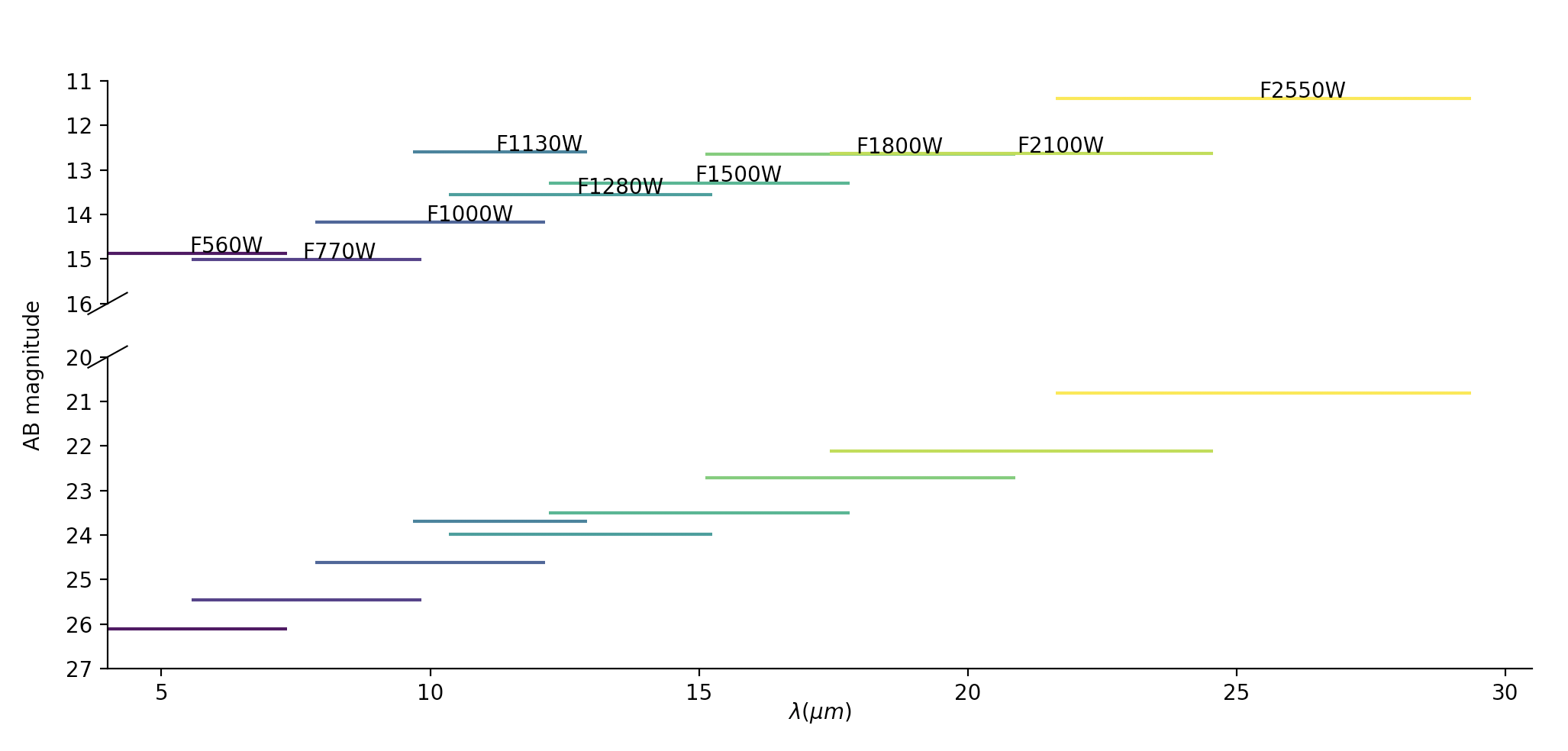}
    
    \caption{MIRI saturation magnitudes in 10\,000 seconds exposure time, and point source sensitivity for 21.4 seconds exposure time. The sizes roughly represent the wavelength range of each filter. There is a break in y-axis (16--20 AB magnitude) to save a large blank space between the sensitivity and saturation limits.}
    \label{fig:sen_bri_MIRI}
\end{figure*}

Figures~\ref{fig:sen_bri_NIRCam} and \ref{fig:sen_bri_MIRI} represent the sensitivity and saturation limits for observing with NIRCam and MIRI filters with a minimum signal-to-noise ratio of 10.

\section{Computational caveats}
\subsection{Computational cost}
\label{app:comp_cost}
Here we address the computational costs of implementing the DeepLIFT from SHAP framework on a photometric data set with a full set of JWST filters and redshift (i.e. 38 features). We compute the time consumption of calculating Shapley values for several sub-samples with different sizes from the training and the validation data sets. We fit an exponential function to the computed time consumption and the corresponding size of the sub-samples as the percentage of the training and validation data sets. We found the following function as the computational cost function for our data set:
\[
\mathcal{Q}(x)= a + (b \times e^{-cx})\enspace{,}
\]
where $a\approx-59.3$, $b\approx51$, and $c\approx-2.06$.
Table~\ref{tab:comp_cost} summarises the computed and estimated computational costs as the time consumption for sub-samples with different sizes. The estimated times are calculated by $\mathcal{Q}(x)$ for a given $x$ as the percentage of the training and validation data sets used to calculate the expectation and Shapley values, respectively. The computed computational costs are derived from an implementation of the algorithm on a MacBook Pro with 9 GHz 6-Core Intel Core i9 processor, and 32 GB 2400 MHz DDR4 memory.

\begin{table}
\caption{Computational costs of calculating Shapley values using DeepLIFT for different sub-samples of validation and training data sets. The instance highlighted in blue is the selected size in this work.  }
\label{tab:comp_cost}
\centering
\begin{tabular}{p{1.5cm}|p{1.3cm}|p{3.1cm}}
\hline \hline
Type & Size (\%) & Time consumption (s)\\
\hline
Computed& 0.1 & 7 \\
Computed& 0.2 & 14.7 \\
Computed& 0.5 & 86.5\\
Computed & 1 & 347\\
\textcolor{blue}{Computed} & $\sim$ 2 &  3041\\
Estimated & 4 & $1.9 \times 10^{5}$\\
Estimated & 5 & $1.5 \times 10^{6}$\\
Estimated & 10 & $4.5 \times 10^{10}$\\
Estimated & 100 & $1.4 \times 10^{91}$\\
\hline

\end{tabular}
\end{table}

\subsection{Reproducibility}
For the sake of reproducibility, we use Keras\citep{chollet2015keras} library from Tensorflow package \citep{tensorflow2015-whitepaper} to build our neural network. We made our code
publicly accessible on GitHub\footnote{\href{https://github.com/ZoeAnsari/InferringSNdustwithNN}{\color{blue}https://github.com/ZoeAnsari/InferringSNdustwithNN}}, for any further evaluation and/or optimisation purposes. However, a specific distribution of computations over the processors has an effect on the training. Therefore, depending on each specific machine in use the reproduced outcome of the neural network can be slightly different \citep[e.g.][]{2021arXiv210203349B}.

\section{S1, S2 feature importance}
\label{app:s1s2}
\begin{figure}
    \centering
    \includegraphics[width=0.45\textwidth]{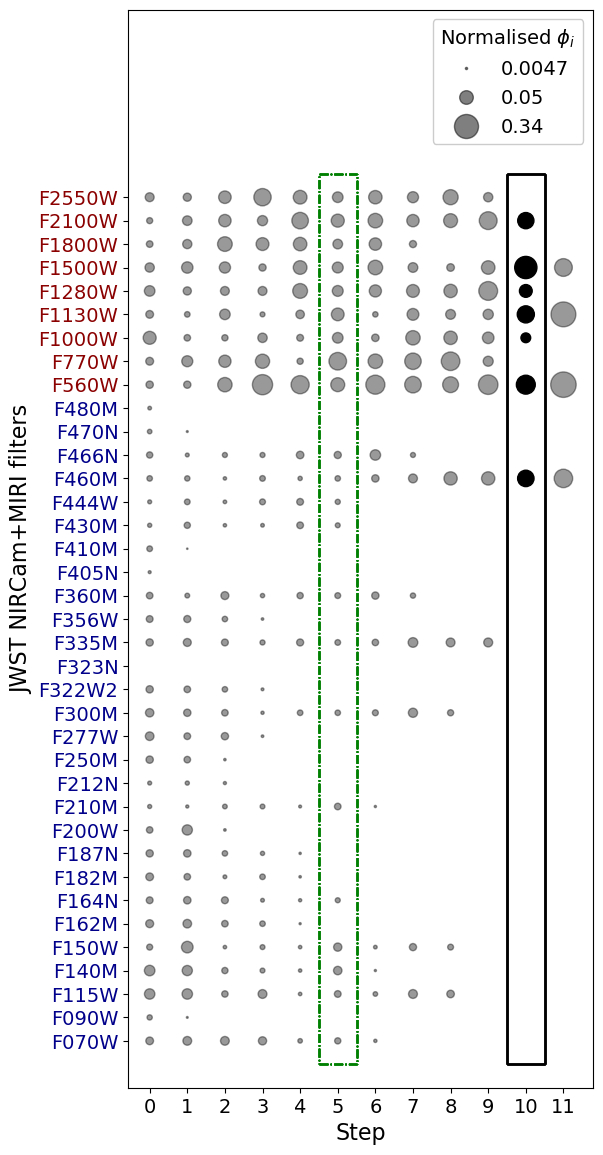}
    \caption{Importance of JWST filters for estimating the amount, temperature and the dust species, in S1. The symbols, relative sizes and colour codes are the same as defined in Figure~\ref{fig:bands_its_s1}.}
    \label{fig:bands_its_s1}
\end{figure}

\begin{figure}
    \centering
    \includegraphics[width=0.45\textwidth]{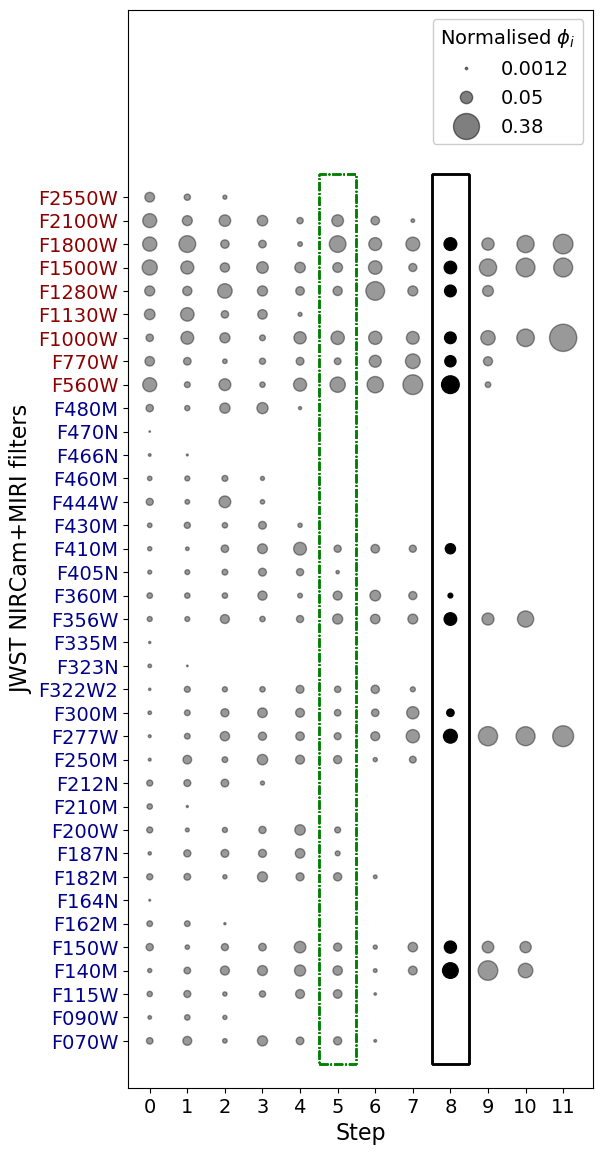}
    \caption{Importance of JWST filters for estimating the amount, temperature and the dust species, in S2. The symbols, relative sizes and colour codes are the same as defined in Figure~\ref{fig:bands_its_s2}.}
    \label{fig:bands_its_s2}
\end{figure}

Figures~\ref{fig:bands_its_s1} and \ref{fig:bands_its_s2} represent the resulting Shapley values obtained for
each step in scenarios S1 and S2.


\section{The effect of optical constants on predicted SEDs}
\label{app:opticalconstants}
\begin{figure*}
    \centering
    \includegraphics[width=\textwidth]{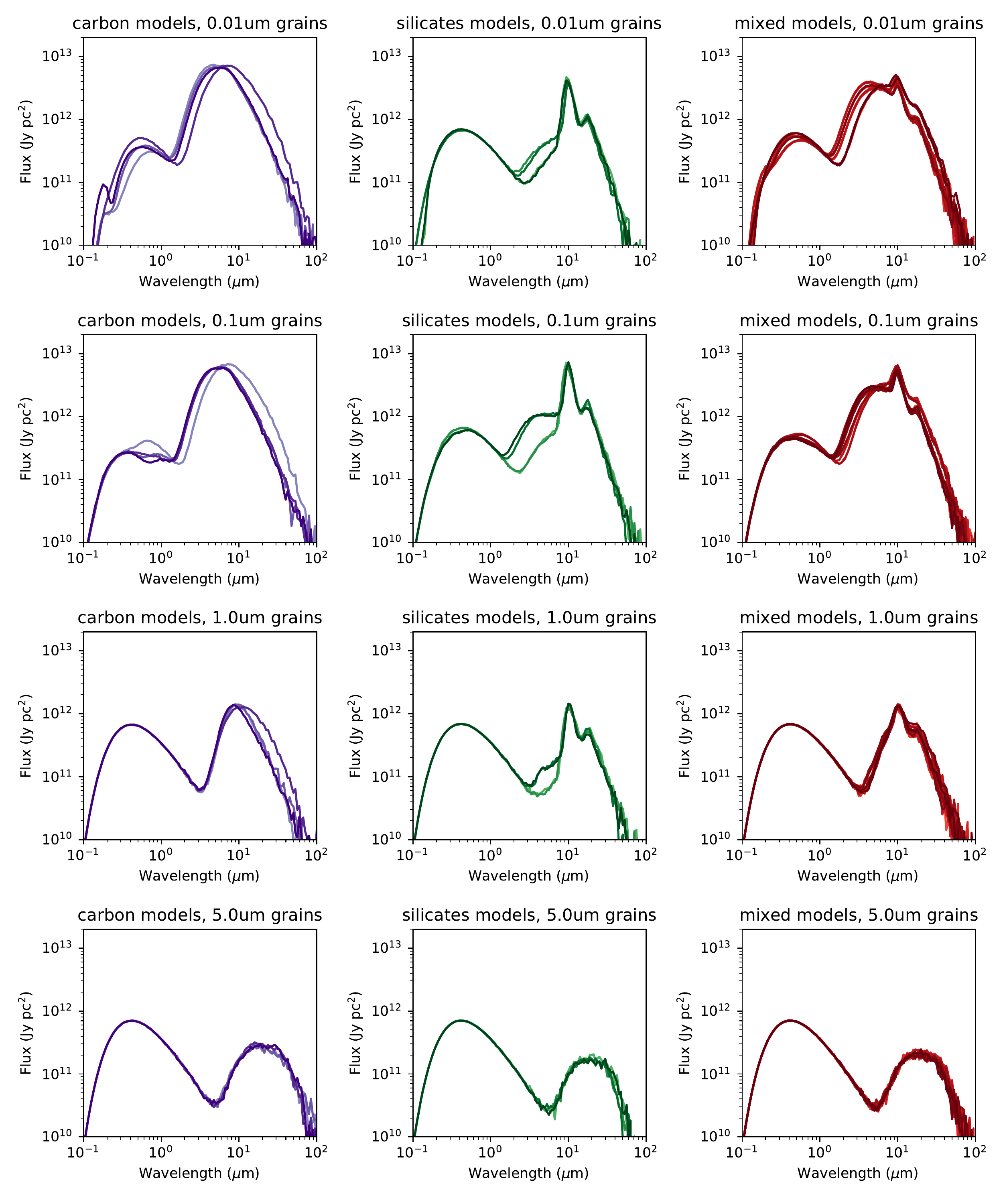}
    \caption{The effect of choice of optical constants on predicted SEDs, for pure and mixed compositions, and grain sizes of 0.01, 0.1, 1.0 and 5.0$\mu$m. SEDs are plotted for all combinations of four sets of carbon and four sets of silicate optical constants, resulting in four SEDS in each panel for pure composition, and 16 for the mixed models.}
    \label{fig:opticalconstantstest_full}
\end{figure*}

As discussed in the text, the choice of optical constants can have a significant effect on the predicted SED. Here, we provide a further illustration of this, using a representative model from our dataset. The model has a mixed composition with 50\% carbon and 50\% silicate grains. Figure~\ref{fig:opticalconstantstest} in the main text shows the predicted SEDs from all combinations of four sets of carbon and four sets of silicate optical constants, for a grain size of 0.1 $\mu$m. In Figure~\ref{fig:opticalconstantstest_full}, we show the sets of predicted SEDs also for grain sizes of 0.01, 1.0 and 5.0 $\mu$m, and for models with the same geometry but pure carbon and pure silicate composition. One can see that the variation between SEDs is largest for smaller grain sizes.

\end{appendix}

%------------------------------------------

\end{document}